\documentclass[longauth]{aa} 

\usepackage{color}
\usepackage{graphicx}
\usepackage{amsmath}
\usepackage{mathrsfs}
\usepackage{txfonts}
\usepackage{natbib,twoopt}
\usepackage{hyperref}
\usepackage[svgnames]{xcolor}
\definecolor{mylightblue}{rgb}{0.19,0.55,0.91}
\hypersetup{colorlinks,breaklinks,
            urlcolor=mylightblue,citecolor=violet,
            anchorcolor=lightgray,linkcolor=mylightblue}
\bibpunct{(}{)}{;}{a}{}{,}             
\makeatletter
  \newcommandtwoopt{\citeads}[3][][]{\href{http://adsabs.harvard.edu/abs/#3}%
    {\def\hyper@linkstart##1##2{}%
     \let\hyper@linkend\@empty\citealp[#1][#2]{#3}}}
  \newcommandtwoopt{\citepads}[3][][]{\href{http://adsabs.harvard.edu/abs/#3}%
    {\def\hyper@linkstart##1##2{}%
     \let\hyper@linkend\@empty\citep[#1][#2]{#3}}}
  \newcommandtwoopt{\citetads}[3][][]{\href{http://adsabs.harvard.edu/abs/#3}%
    {\def\hyper@linkstart##1##2{}%
     \let\hyper@linkend\@empty\citet[#1][#2]{#3}}}
  \newcommandtwoopt{\citeyearads}[3][][]%
    {\href{http://adsabs.harvard.edu/abs/#3}
    {\def\hyper@linkstart##1##2{}%
     \let\hyper@linkend\@empty\citeyear[#1][#2]{#3}}}
\makeatother
\DeclareSymbolFont{mymath}{T1}{ybv}{m}{it}
\SetSymbolFont{mymath}{normal}{T1}{ybv}{m}{it}
\DeclareSymbolFontAlphabet{\mathnormal}{mymath}
\DeclareMathSymbol{\vel}{\mathalpha}{mymath}{`v}

\begin{document} 

   \title{The \emph{Gaia}-ESO Survey: Galactic evolution of lithium from 
     iDR6\thanks{Based on data products from observations made with ESO 
       Telescopes at the La Silla Paranal Observatory under programmes 
       188.B-3002, 193.B-0936, and 197.B-1074.}$^,$\thanks{Full 
       Tables~\ref{tab:FLDsample} and \ref{tab:ORBparams} are only available at 
       the CDS via anonymous ftp to \url{cdsarc.u-strasbg.fr (130.79.128.5)} 
       or via \url{http://cdsarc.u-strasbg.fr/viz-bin/cat/J/A+A/xxx/Ayyy}}}


   \author{D.~Romano
          \inst{1},
          L.~Magrini
          \inst{2},
          S.~Randich
          \inst{2},
          G.~Casali
          \inst{2, 3},
          P.~Bonifacio
          \inst{4},
          R.~D.~Jeffries
          \inst{5},
          F.~Matteucci
          \inst{6, 7, 8},
          E.~Franciosini
          \inst{2},
          L.~Spina
          \inst{9},
          G.~Guiglion
          \inst{10},
          C.~Chiappini
          \inst{10},
          A.~Mucciarelli
          \inst{11, 1},
          P.~Ventura
          \inst{12},
          V.~Grisoni
          \inst{13, 7},
          M.~Bellazzini
          \inst{1},
          T.~Bensby
          \inst{14},
          A.~Bragaglia
          \inst{1},
          P.~de Laverny
          \inst{15},
          A.~J.~Korn
          \inst{16},
          S.~L.~Martell
          \inst{17, 18},
          G.~Tautvai{\v s}ien{\. e}
          \inst{19},
          G.~Carraro
          \inst{20},
          A.~Gonneau
          \inst{21},
          P.~Jofr{\'e}
          \inst{22},
          E.~Pancino
          \inst{2, 23},
          R.~Smiljanic
          \inst{24},
          A.~Vallenari
          \inst{9},
          X.~Fu
          \inst{25},
          M.~L.~Guti{\'e}rrez Albarr{\'a}n
          \inst{26},
          F.~M.~Jim{\'e}nez-Esteban
          \inst{27},
          D.~Montes
          \inst{26},
          F.~Damiani
          \inst{28},
          M.~Bergemann
          \inst{29},
          \and
          C.~Worley
          \inst{21}
          }

   \institute{INAF, Osservatorio di Astrofisica e Scienza dello Spazio, 
     Via Gobetti 93/3, 40129 Bologna, Italy\\
     \email{donatella.romano@inaf.it}
     \and INAF, Osservatorio Astrofisico di Arcetri, Largo E. Fermi 5, 50125 
     Firenze, Italy
     \and Dipartimento di Fisica e Astronomia, Universit\`a degli Studi di 
     Firenze, Via G. Sansone 1, 50019 Sesto Fiorentino, Firenze, Italy
     \and GEPI, Observatoire de Paris, Universit{\'e} PSL, CNRS, Place Jules
     Janssen, 92190 Meudon, France
     \and Astrophysics Group, Keele University, Keele, Staffordshire ST5 5BG, 
     UK
     \and Dipartimento di Fisica, Sezione di Astronomia, 
     Universit\`a di Trieste, Via Tiepolo 11, 34131 Trieste, Italy
     \and INAF, Osservatorio Astronomico di Trieste, 
     Via Tiepolo 11, 34131 Trieste, Italy
     \and INFN, Sezione di Trieste, Via Valerio 2, 34127 Trieste, Italy
     \and INAF, Osservatorio Astronomico di Padova, Vicolo dell'Osservatorio 5, 
     35122 Padova, Italy
     \and Leibniz-Institut f{\"u}r Astrophysik Potsdam (AIP), An der
     Sternwarte 16, 14482 Potsdam, Germany
     \and Dipartimento di Fisica e Astronomia, Universit\`a degli Studi di 
     Bologna, Via Gobetti 93/2, 40129 Bologna, Italy
     \and INAF, Osservatorio Astronomico di Roma, Via Frascati 33, 00077 Monte 
     Porzio Catone, Roma, Italy
     \and SISSA, Via Bonomea 265, 34136 Trieste, Italy
     \and Lund Observatory, Department of Astronomy and Theoretical Physics, 
     Box 43, 221 00 Lund, Sweden
     \and Universit{\'e} C{\^o}te d'Azur, Observatoire de la C{\^o}te d'Azur, 
     CNRS, Laboratoire Lagrange, Bd de l’Observatoire, CS 34229, 06304 Nice 
     Cedex 4, France
     \and Observational Astrophysics, Department of Physics and Astronomy, 
     Uppsala University, Box 516, 751 20 Uppsala, Sweden
     \and School of Physics, UNSW, Sydney, NSW 2052, Australia
     \and Centre of Excellence for All-Sky Astrophysics in Three Dimensions
     (ASTRO 3D), Australia
     \and Institute of Theoretical Physics and Astronomy, Vilnius University,
     Sauletekio av.~3, 10257 Vilnius, Lithuania
     \and Dipartimento di Fisica e Astronomia, Universit{\`a} di Padova, Vicolo 
     dell'Osservatorio 3, 35122 Padova, Italy
     \and Institute of Astronomy, University of Cambridge, Madingley Road, 
     Cambridge CB3 0HA, UK
     \and N{\'u}cleo de Astronom{\'i}a, Facultad de Ingenier{\'i}a,
     Universidad Diego Portales, Av. Ej{\'e}rcito 441, Santiago, Chile
     \and Space Science Data Center-ASI, Via del Politecnico SNC, 00133 
     Roma, Italy
     \and Nicolaus Copernicus Astronomical Center, Polish Academy of Sciences, 
     ul. Bartycka 18, 00-716 Warsaw, Poland
     \and The Kavli Institute for Astronomy and Astrophysics at Peking 
     University, 100871 Beijing, China
     \and Departamento de F{\'i}sica de la Tierra y Astrof{\'i}sica and 
     IPARCOSUCM, Instituto de F{\'i}sica de Part{\'i}culas y del Cosmos de la 
     UCM, Facultad de Ciencias F{\'i}sicas, Universidad Complutense de Madrid,
     28040 Madrid, Spain
     \and Departmento de Astrof{\'i}sica, Centro de Astrobiolog{\' i}a
     (INTA-CSIC), ESAC Campus, Camino Bajo del Castillo s/n, 28692 Villanueva
     de la Ca{\~n}ada, Madrid, Spain
     \and INAF, Osservatorio Astronomico di Palermo, Piazza del Parlamento 1, 
     90134 Palermo, Italy
     \and Max-Planck Institut f{\"u}r Astronomie, K{\"o}nigstuhl 17, 69117 
     Heidelberg, Germany
     \\
             }

   \date{Received 19 May 2021 / Accepted 18 June 2021}

   \titlerunning{Galactic evolution of lithium from GES iDR6}
   \authorrunning{Romano et al.}

   \abstract
      {After more than 50 years, astronomical research still struggles to 
        reconstruct the history of lithium enrichment in the Galaxy and to 
        establish the relative importance of the various \element[ ][7]{Li} 
        sources in enriching the interstellar medium (ISM) with this fragile 
        element.}
      {To better trace the evolution of lithium in the Milky Way discs, we 
        exploit the unique characteristics of a sample of open clusters (OCs) 
        and field stars for which high-precision \element[ ][7]{Li} abundances 
        and stellar parameters are homogeneously derived by the {\it Gaia}-ESO 
        Survey (GES).}
      {We derive possibly un-depleted \element[ ][7]{Li} abundances for 26 OCs 
        and star forming regions with ages from young ($\sim$3~Myr) to old 
        ($\sim$4.5~Gyr), spanning a large range of galactocentric distances, $5 
        < R_{\mathrm{GC}}$/kpc~$< 15$, which allows us to reconstruct the local 
        late Galactic evolution of lithium as well as its current abundance 
        gradient along the disc. Field stars are added to look further back in 
        time and to constrain \element[ ][7]{Li} evolution in other Galactic 
        components. The data are then compared to theoretical tracks from 
        chemical evolution models that implement different \element[ ][7]{Li} 
        forges.}
      {Thanks to the homogeneity of the GES analysis, we can combine the 
        maximum average \element[ ][7]{Li} abundances derived for the clusters 
        with \element[ ][7]{Li} measurements in field stars. We find that the 
        upper envelope of the \element[ ][7]{Li} abundances measured in field 
        stars of nearly solar metallicities ($-0.3 <$~[Fe/H]/dex~$< +0.3$) 
        traces very well the level of lithium enrichment attained by the ISM as 
        inferred from observations of cluster stars in the same metallicity 
        range. We confirm previous findings that the abundance of 
        \element[ ][7]{Li} in the solar neighbourhood does not decrease at 
        super-solar metallicity. The comparison of the data with the chemical 
        evolution model predictions favours a scenario in which the majority of 
        the \element[ ][7]{Li} abundance in meteorites comes from novae. 
        Current data also seem to suggest that the nova rate flattens out at 
        later times. This requirement might have implications for the masses of 
        the white dwarf nova progenitors and deserves further investigation. 
        Neutrino-induced reactions taking place in core-collapse supernovae 
        also produce some fresh lithium. This likely makes a negligible 
        contribution to the meteoritic abundance, but could be responsible for 
        a mild increase in the \element[ ][7]{Li} abundance in the ISM of 
        low-metallicity systems that would counterbalance the astration 
        processes.}
      {}

   \maketitle
%

   \section{Introduction}
   \label{sec:intro}

   Of all the elements in the periodic table, the main isotope of lithium, 
   \element[ ][7]{Li}, has without any doubt the most complex origin and 
   evolution. Its nuclei were produced in significant amounts during Big Bang 
   nucleosynthesis (BBN), 10 per cent directly as \element[ ][7]{Li} in the 
   first few minutes and 90 per cent as \element[ ][7]{Be} that decayed to 
   \element[ ][7]{Li} later on \citep{2007ARNPS..57..463S,2011AstL...37..367K,
     2020JCAP...03..010F}. Thereafter, both spallation processes that take 
   place in the interstellar medium (ISM) and nuclear burning in stars 
   contribute to increase its abundance from the primordial value to that 
   currently observed in meteorites and young T\,Tauri stars.

   It is a little disconcerting that none of the proposed \element[ ][7]{Li} 
   production channels have been firmly assessed yet. The factor of three 
   difference between the primordial abundance of lithium predicted by the 
   standard BBN (SBBN) theory assuming the {\it Planck} baryon density, 
   $A$(Li)$_{\mathrm{P}}^{\mathrm{th}} \sim$~2.7 dex \citep{2018PhR...754....1P}, 
   and that inferred from observations of halo stars on the `Spite plateau' 
   that should not have depleted their \element[ ][7]{Li}, 
   $A$(Li)$_{\mathrm{P}}^{\mathrm{obs}} \sim$~2.2 dex \citep{1982A&A...115..357S,
     1997MNRAS.285..847B,2010A&A...522A..26S}, constitutes the well-known 
   cosmological lithium problem and raises the question as to whether the BBN 
   model \citep[e.g.][]{2012ApJ...744..158C,2016PhRvL.116u1303G,
     2017ApJ...834..165H,2019ApJ...872..172L,2020A&A...633L..11C} or stellar 
   physics \citep[][and references therein]{1984ApJ...282..206M,
     2005A&A...442..961C,2012MNRAS.419.2195M,2012ApJ...753...48N,
     2015MNRAS.452.3256F,2020MNRAS.497L..30G} is to blame.

   As \element[ ][7]{Li} enters the chemical composition of successive stellar 
   generations, it is swiftly burned through \element[ ][7]{Li}(p, 
   $\alpha$)\element[ ][4]{He} reactions already on the pre-main sequence 
   wherever the temperature exceeds about $2.5 \times 10^6$~K. For 
   \element[ ][7]{Li} to be produced in stars, it is necessary that 
   \element[ ][7]{Be} be formed first, at temperatures exceeding $4 \times 
   10^7$~K. The newly formed \element[ ][7]{Be} must then be promptly carried 
   by convection to cooler layers, where it decays to \element[ ][7]{Li} that 
   is preserved and, eventually, ejected. This mechanism was first proposed by 
   \citet{1955ApJ...121..144C} and \citet{1971ApJ...164..111C} to explain the 
   existence of Li-rich asymptotic giant branch (AGB) stars. Low-mass stars on 
   the first ascent of the red giant branch (RGB) may also show enhanced 
   lithium if some extra deep mixing is associated with the cool bottom 
   processing \citep{1999ApJ...510..217S}. Although some Li-rich and very 
   Li-rich giant stars do exist \citep{1982ApJ...255..577W,2011ApJ...743..107R,
     2014ApJ...784L..16S,2016ApJ...819..135K,2018A&A...617A...4S,
     2019ApJ...878L..21S,2020A&A...633A..34C,2020MNRAS.494.1348D}, low- and 
   intermediate-mass stars are unlikely to contribute significantly to the 
   \element[ ][7]{Li} enrichment on galactic scales \citep{2000A&A...363..605V,
     2020A&A...641A.103V} unless some extreme assumptions are made on the 
   effectiveness of mass loss along the giant branches that should be maximal 
   exactly when the stars are Li-rich \citep{2001A&A...374..646R,
     2001ApJ...559..909T}.

   Things change if low- and intermediate-mass stars are hosted in cataclysmic 
   variables. Classical nova explosions occur in these binary systems when the 
   white dwarf (WD) primary accretes matter from a secondary star that 
   overflows its Roche lobe. Once the temperature of the deepest layers 
   accreted on top of the WD exceeds $\sim$$7 \times 10^7$~K, a thermonuclear 
   runaway is initiated, which leads to an outburst that does not disrupt the 
   WD. After the ejection of the envelope, which is enriched in rare isotopes, 
   the process is re-initiated \citep[][and references 
     therein]{2007JPhG...34..431J}; on average, the typical nova experiences 
   $10^4$ outbursts during its lifetime \citep{1978MNRAS.183..515B}.

   Although theoretical calculations of \element[ ][7]{Li} production during 
   thermonuclear runaways demonstrated long ago that novae could be important 
   contributors to the Galactic \element[ ][7]{Li} enrichment 
   \citep{1978ApJ...222..600S,1991A&A...248...62D}, until very recently there 
   was no observational evidence to support the theory. 
   \citet{2015ApJ...808L..14I} and \citet{2015Natur.518..381T} first detected, 
   respectively, the blueshifted \ion{Li}{I}\,$\lambda$\,6708\,$\AA$ line in 
   the early spectra of nova V1369\,Cen and strong \ion{Be}{II} features in the 
   spectra of nova V339\,Del. As \element[ ][7]{Be} was detected in more 
   objects \citep[e.g.][]{2018MNRAS.481.2261S,2020MNRAS.492.4975M}, a 
   discrepancy between the observational abundance estimates and theoretical 
   predictions of nucleosynthesis models became apparent, with the former 
   exceeding the latter by up to one order of magnitude. The reason for this 
   may reside in the (wrong) assumption of equality between the relative 
   fractions of the ions \ion{Be}{II}/Be and \ion{Ca}{II}/Ca in the nova ejecta 
   \citep{2020AstL...46...92C} and/or in an overly simplistic treatment of the 
   complex structure of the ejecta itself \citep{2020A&A...639L..12S}. However, 
   \citet{2021MNRAS.501L..33D} show that an enhanced abundance of 
   \element[ ][4]{He} in nova envelopes would push the \element[ ][7]{Be} 
   abundance predicted by nova models closer to the observed values. From the 
   above it seems evident that at present neither spectroscopic observations 
   nor stellar models can provide us with a sound, quantitative estimate of 
   \element[ ][7]{Li} production from nova systems.

   The flux of neutrinos that emerges from the collapsing core of an exploding 
   massive star is another trigger of \element[ ][7]{Li} synthesis 
   \citep[see][for the first realistic exploration of the $\nu$-process in 
     stars]{1990ApJ...356..272W}. \citet{2018ApJ...865..143S} have presented 
   results for a grid of solar-metallicity models with stellar masses in the 
   range 13--30~M$_\sun$. Their models take into account up-to-date 
   cross-sections for the neutrino-induced reactions and a more realistic 
   treatment of the neutrino transport in supernova (SN) simulations that 
   results in lower-energy spectra for all neutrino families. Due to the 
   reduction in neutrino energies, the efficiency of the $\nu$-process is 
   diminished. Their \element[ ][7]{Li} yields range from $1 \times 
   10^{-8}$~M$_\sun$ to $2 \times 10^{-8}$~M$_\sun$, one order of magnitude lower 
   than the older estimates by \citet{1995ApJS..101..181W}. However, the use of 
   time-dependent neutrino emission spectra and neutrino oscillation and 
   metallicity effects are expected to affect \element[ ][7]{Li} production in 
   the $\nu$-process, and larger yields could be obtained 
   \citep{2019ApJ...872..164K,2019ApJ...876..151S}. Spectroscopic observations 
   of Li in SN remnants could establish the mechanism of \element[ ][7]{Li} 
   production in massive stars. The only measurement of this kind we are aware 
   of \citep{2012ApJ...750L..15T} finds no evidence of \element[ ][7]{Li} 
   synthesis by neutrino-induced spallation in core-collapse SNe. However, the 
   absence of evidence is not evidence of absence; clearly, we need to probe 
   many more diffuse molecular clouds near other SN remnants.

   We conclude this roundup of \element[ ][7]{Li} factories by briefly 
   reviewing another triplet of possible sources. First, high-energy Galactic 
   cosmic ray (GCR) nuclei impacting stationary ISM atoms have been known for 
   50 years as \element[ ][7]{Li} sources \citep{1970Natur.226..727R,
     1971A&A....15..337M}. However, they produce no more than 20--25 per cent 
   of the meteoritic \element[ ][7]{Li} abundance 
   \citep[see][]{2001A&A...374..646R,2012A&A...542A..67P}. Second, spallation 
   in hot advection-dominated accretion flows can make quiescent soft X-ray 
   transients quite efficient at producing \element[ ][7]{Li} \citep[][and 
     references therein]{1997ApJ...486..363Y,2008ApJ...673L..51F}. Yet, under 
   reasonable hypotheses, the contribution from these objects to 
   \element[ ][7]{Li} pollution on galactic scales turns out to be negligible 
   \citep{2001A&A...374..646R}. A similar conclusion is reached regarding the 
   creation of lithium in flares of active stars \citep[][and references 
     therein]{2020RAA....20..104K}.

   The Galactic chemical evolution (GCE) models adopted in this work take all 
   of the \element[ ][7]{Li} sources mentioned above, bar the last two, into 
   account. The models predict the amount of lithium a star will be formed with 
   at any given time for different Galactic components. The stellar abundance 
   then falls over time, which makes the upper envelope of the lithium 
   abundances measured in warm, dwarf stars the useful thing to measure (see 
   below). The trends of $A$(Li) as functions of [Fe/H] and age predicted for 
   solar neighbourhood stars are compared to spectroscopic data from the sixth 
   internal data release (iDR6) of the {\it Gaia}-ESO Spectroscopic 
   Survey\footnote{\url{https://www.gaia-eso.eu}} 
   \citep[GES;][]{2012Msngr.147...25G,2013Msngr.154...47R}. Moreover, we 
   compare, for the first time, the theoretical $A$(Li) gradient to 
   \element[ ][7]{Li} abundances measured in  objects covering a large range of 
   galactocentric distances. After \citet{1988A&A...192..192R}, it became 
   common practice to assume that the upper envelope of the abundances of 
   \element[ ][7]{Li} measured in warm, dwarf Galactic stars tracks the 
   \element[ ][7]{Li} enrichment of the ISM fairly faithfully. In more recent 
   years, the mean values of the highest \element[ ][7]{Li} abundances in bins 
   of metallicity have been put forward as a more appropriate indicator 
   \citep[e.g.][]{2004MNRAS.349..757L,2015A&A...576A..69D,2016A&A...595A..18G,
     2018A&A...610A..38F}. It has also become clear that the mixing of 
   different populations can lead to spurious trends; in particular, the 
   effects of stellar radial migration must be taken into account 
   \citep{2019A&A...623A..99G}. Finally, it has been shown that the trends of 
   \element[ ][7]{Li} with metallicity, age, and galactocentric distance are 
   best traced when considering cluster stars that have not undergone any 
   depletion \citep[see][]{2020A&A...640L...1R}.

   This paper is organised as follows. Section~\ref{sec:data} details the 
   sample selection, \element[ ][7]{Li} abundance determination, and the 
   kinematical and dynamical analysis. We introduce the adopted GCE model in 
   Sect.~\ref{sec:model}. Section~\ref{sec:results} describes the evolution of 
   \element[ ][7]{Li} in the thick and thin discs emerging from the field and 
   cluster star samples selected from GES iDR6. Lithium observations are 
   compared to the predictions of chemical evolution models in 
   Sect.~\ref{sec:results} and further discussed in Sect.~\ref{sec:discuss}. 
   Finally, in Sect.~\ref{sec:end} we draw our conclusions.

   \section{Dataset}
   \label{sec:data}

\begin{table*}
\caption{Stellar parameters and lithium abundances of 3210 stars in the field, 
  ordered by increasing age.}
\label{tab:FLDsample}
\centering
\begin{scriptsize}
\begin{tabular}{l c c c c c c c c c c c c c c c}
\hline\hline
GES name         & RA        & Dec         & $T_{\mathrm{eff}}$ & $\delta T_{\mathrm{eff}}$ & $\log g$ & $\delta \log g$ & [Fe/H]  & $\delta$[Fe/H] & $A$(Li)   & $\delta A$(Li) & [$\alpha$/Fe] & $\delta [\alpha$/Fe] & Age    & $\delta$Age & $R_{\mathrm{GC}}$ \\
                 & [deg]     & [deg]       & [K]             & [K]                     & [dex]    & [dex]           & [dex]   & [dex]           & [dex]      & [dex]          & [dex]         & [dex]                & [Gyr]  & [Gyr]       & [pc] \\
\hline
17530917-2937418 & 268.28821 & $-$29.6283  & 6807           & 66                     & 4.30     & 0.17            & 0.18    & 0.06            & $<$\,2.62 &  --            & --            & --                   & 1.06  & 0.56       & 6009 \\
17524903-2926366 & 268.20429 & $-$29.4435 & 6754           & 67                     & 4.45     & 0.18            & 0.08    & 0.05            & $<$\,2.28 &  --            & --            & --                   & 1.13  & 0.61       & 5916 \\ 
10371153-5838408 & 159.29804 & $-$58.6447 & 6721           & 67                     & 4.23     & 0.17            & 0.24    & 0.05            & 3.36      & 0.06           & --            & --                   & 1.18  & 0.60       & 7683 \\
17530416-2936240 & 268.26733 & $-$29.6067 & 6887           & 65                     & 4.24     & 0.17            & 0.04    & 0.05            & 2.49      & 0.19           & --            & --                   & 1.19  & 0.60       & 4419 \\
17525280-2931432 & 268.22000 & $-$29.5287 & 6785           & 66                     & 4.25     & 0.18            & 0.15    & 0.05            & 2.97      & 0.06           & --            & --                   & 1.19  & 0.61       & 6164 \\
\ldots           & \ldots    & \ldots     & \ldots         & \ldots                 & \ldots   & \ldots          & \ldots  & \dots           & \ldots   & \ldots         & \ldots       & \ldots               & \ldots & \ldots      & \ldots \\
\hline
\end{tabular}
\end{scriptsize}
\tablefoot{
The reported errors are the result of the GES homogenisation procedure and 
include random and systematic error sources, apart from the uncertainties on 
\element[ ][7]{Li} abundances (see text) and on age determinations (which were 
computed as the average of the half widths of the 68\% confidence intervals 
calculated by aussieq2). The table is available in its entirety at the CDS. A 
portion is shown here for guidance regarding its form and contents.
}
\end{table*}

   This study is based on the last internal data release of the GES. We take 
   advantage of the \element[ ][7]{Li} abundances and stellar parameters 
   homogeneously determined for pre-main-sequence, main-sequence, turn-off, and 
   sub-giant stars observed with the multi-object optical fibre facility FLAMES 
   \citep[Fibre Large Array Multi Element Spectrograph;][]{2002Msngr.110....1P} 
   in the Milky Way field and in open clusters (OCs). In particular, regarding 
   the determination of \element[ ][7]{Li}, OC stars were observed either with 
   the Ultraviolet and Visual Echelle Spectrograph (UVES\footnote{See 
     \citet{2014A&A...570A.122S} and \citet{2015A&A...576A..80L} for a complete 
     description of the analysis of the FLAMES-UVES spectra of, respectively, 
     FGK and pre-main-sequence stars in the GES.}) and the 580 setup or with 
   GIRAFFE and the HR15N setup; both configurations include the lithium lines. 
   The HR15N grating was not employed for observations of stars in GES field 
   samples; however, targets observed in cluster fields that were later 
   recognised as non-members may have their \element[ ][7]{Li} abundances 
   derived from GIRAFFE spectra. Our GES iDR6 star sample was additionally 
   cross-matched with the {\it Gaia} Early Data Release 3 (EDR3) catalogue 
   \citep{2016A&A...595A...1G,2020arXiv201201533G} to perform a full 
   chemo-dynamical characterisation of the stars.

   \subsection{Lithium abundance determination and sample selection}
   \label{sec:Liabund}

   One-dimensional (1D) local thermodynamical equilibrium (LTE) abundances of 
   \element[ ][7]{Li} in GES iDR6 were obtained from equivalent width 
   measurements of the \ion{Li}{I}\,$\lambda$\,6708\,$\AA$ spectral feature. 
   The abundance derivation procedure (Franciosini et al., in prep.) involved 
   the use of a new set of curves of growth specifically defined for the GES 
   that relies on the grid of synthetic spectra described in \citet[][see also 
     \citealt{2012A&A...544A.126D}]{2016A&A...595A..18G}; for GIRAFFE spectra, 
   where the lithium feature is blended with the nearby 
   \ion{Fe}{I}\,$\lambda$\,6707.4\,$\AA$ line, a consistent correction was 
   applied. Upper limits were derived when the line was not detectable; more 
   specifically, the larger of the measured equivalent width and the associated 
   uncertainty was assigned as upper limit when the line was not detected or 
   barely visible.

   We did not correct the \element[ ][7]{Li} abundances for 3D non-LTE effects, 
   since they are almost negligible for the majority of the stars in our 
   sample, barely reaching $-$0.1~dex for a handful of stars \citep[see][their 
     Figs.~1 and 2]{2021arXiv210504866M}. According to recent work by 
   \citet{2021MNRAS.500.2159W}, a systematic $-$0.1~dex correction applies to 
   metal-rich dwarfs with $T_{\mathrm{eff}}$ between 5000 and 6500~K. The 
   correction then decreases to zero between 6500 and 7000~K. For the coolest, 
   metal-rich pre-main-sequence stars in clusters the corrections may be as 
   severe as $-0.3$~dex, but these values are very uncertain, since they rely 
   on extrapolations of extant grids.

   The field stars used in the present paper were selected as: (i) field stars 
   that are non-members of the old and intermediate-age OCs (age~$>$ 120~Myr), 
   taking into account all stars not selected as member stars on the basis of 
   their radial velocities from the GES, proper motions, and parallaxes from 
   {\it Gaia} EDR3, and (ii) stars observed in the GES field samples, as 
   indicated by the keywords  {\it GES$\_$MW}, {\it GES$\_$MW$\_$BL}, {\it 
     GES$\_$K2}, {\it GES$\_$CR} in the field {\it GES$\_$FLD}. We combined the 
   two samples, applying a further selection on stellar parameter uncertainties 
   ($\delta\,T_{\mathrm{eff}} < 100$~K, $\delta\,\log g < 0.2$~dex, 
   $\delta\,$[Fe/H] $< 0.15$~dex) and including both stars with measured 
   lithium abundances with error on $A$(Li) lower than 0.25~dex and upper 
   limits. We cross-matched our catalogue with {\it Gaia} EDR3 and computed the 
   stellar luminosities using the geometric distances from 
   \citet{2021AJ....161..147B} and the $G$ magnitudes of {\it Gaia}, converted 
   into $V$ magnitudes using the colours $G_\mathrm{BP}$ and $G_\mathrm{RP}$, as 
   stated in \citet{2021arXiv210504866M}. To compute the bolometric magnitudes 
   we used the bolometric corrections BC($K$) tabulated by \citet[][their 
     Table~5]{1999A&AS..140..261A}, which are based on $V-K$ colours, and $K$ 
   magnitudes from 2MASS \citep{2006AJ....131.1163S}.

   After removing the giant stars from the sample, we were left with 6207 
   late-type stars with effective temperatures ($T_{\mathrm{eff}}$) in the range 
   5300--7000~K, surface gravities ($\log g$) of 3.5 to 4.6 (in cgs units), and 
   metallicities $-1.5 <$~[Fe/H]/dex~$< +$0.5, which were observed with either 
   the FLAMES-UVES 580 setup at high resolution ($R \simeq 47\,000$) or the 
   FLAMES-GIRAFFE HR15N setup at medium resolution ($R \simeq 19\,000$). To 
   further improve the quality of the sample, we retained only the stars with 
   the highest-quality spectra (i.e. signal-to-noise ratio S/N~$\ge 50$). This 
   almost halved our working sample, which finally consists of 3210 stars. 
   Stars with $\log g \simeq 3.5$~dex may undergo the first dredge up and, 
   thus, display lithium abundances lower than the initial value. Yet, our 
   sample contains only 7 stars with $3.5 < \log g < 3.7$, which makes the 
   exact $\log g$ cut rather irrelevant. 

\begin{table*}
\caption{Parameters and average maximum lithium abundance for the selected 
  clusters, sorted by increasing age.}
\label{tab:OCsample}
\centering
\begin{footnotesize}
\begin{tabular}{l c r c c c c l}
\hline\hline
Cluster                      & Age   & $R_{\mathrm{GC}}$ & [Fe/H]           & $A$(Li)$_\mathrm{max}$ & \# of stars & $T_{\mathrm{eff}}$ range \\
                             & [Gyr] & [kpc]          & [dex]            & [dex]                 &             & [K]                    \\
\hline
$\rho$\,Oph\tablefootmark{a} & 0.003 &  7.88          & ($-0.265$)       & $3.28 \pm 0.13$       & 25          & 3006--4536 (pre-main sequence) \\
Alessi\,43                   & 0.011 &  8.18          & $+0.02 \pm 0.06$ & $3.27 \pm 0.22$       & 46          & 3513--5135 (pre-main sequence) \\
25\,Ori\tablefootmark{b}     & 0.013 &  8.31          & ($0.02$)         & $3.18 \pm 0.14$       & 7           & 3119--3386 (pre-main sequence) \\
Collinder\,197               & 0.014 &  8.20          & ($0.02$)         & $3.21 \pm 0.15$       & 64          & 3515--4441 (pre-main sequence) \\
NGC\,2232                    & 0.018 &  8.27          & ($0.005$)        & $3.22 \pm 0.11$       & 14          & 5031--6891 (young) \\
NGC\,2547                    & 0.032 &  8.05          & ($-0.055$)       & $3.35 \pm 0.10$       & 8           & 5748--6286 (young) \\
IC\,4665                     & 0.033 &  7.71          & ($0.00$)         & $3.44 \pm 0.10$       & 5           & 5674--6133 (young) \\
NGC\,6405                    & 0.035 &  7.54          & ($-0.01$)        & $3.31 \pm 0.08$       & 7           & 5848--6419 \\
IC\,2602                     & 0.036 &  7.95          & ($-0.01$)        & $3.31 \pm 0.13$       & 4           & 5766--6438 (young) \\
Blanco\,1                    & 0.105 &  7.96          & $-0.12 \pm 0.07$ & $3.15 \pm 0.04$       & 7           & 5923--6476 (young) \\
NGC\,6067                    & 0.126 &  6.16          & $+0.03 \pm 0.05$ & $3.38 \pm 0.09$       & 15          & 6518--8000 \\
NGC\,6709                    & 0.190 &  7.22          & $-0.03 \pm 0.06$ & $3.31 \pm 0.01$       & 2           & 6628--7410 \\
NGC\,2516                    & 0.240 &  7.98          & $-0.04 \pm 0.05$ & $3.33 \pm 0.07$       & 3           & 6531--6839 \\
Berkeley\,30                 & 0.295 & 13.59          & $-0.15 \pm 0.05$ & $3.11 \pm 0.17$       & 14          & 6630--6980 \\
NGC\,6705                    & 0.309 &  6.02          & $+0.02 \pm 0.01$ & $3.34 \pm 0.02$       & 7           & 6673--6984 \\
NGC\,3532                    & 0.398 &  7.85          & $-0.01 \pm 0.05$ & $3.25 \pm 0.07$       & 4           & 6825--6865 \\
NGC\,6802                    & 0.660 &  6.71          & $+0.14 \pm 0.04$ & $3.32 \pm 0.02$       & 2           & 6677--6919 \\
NGC\,2355                    & 1.000 &  9.84          & $-0.07 \pm 0.05$ & $3.17 \pm 0.09$       & 6           & 6716--6975 \\
Berkeley\,81                 & 1.148 &  5.21          & $+0.22 \pm 0.05$ & $3.39 \pm 0.11$       & 1           & 6836 \\
Berkeley\,73                 & 1.413 & 14.97          & $-0.26 \pm 0.05$ & $3.16 \pm 0.11$       & 8           & 6721--6885 \\
Berkeley\,44                 & 1.445 &  6.58          & $+0.22 \pm 0.05$ & $3.21 \pm 0.10$       & 9           & 6564--6761 \\
NGC\,2158                    & 1.549 & 13.18          & $-0.16 \pm 0.05$ & $3.19 \pm 0.09$       & 5           & 6585--6889 \\
Ruprecht\,134                & 1.660 &  5.44          & $+0.27 \pm 0.05$ & $3.47 \pm 0.04$       & 8           & 6697--6879 \\
NGC\,2420                    & 1.738 & 10.49          & $-0.16 \pm 0.05$ & $3.19 \pm 0.07$       & 22          & 6378--6769 (turn-off) \\
Trumpler\,20                 & 1.862 &  6.82          & $+0.13 \pm 0.04$ & $3.29 \pm 0.12$       & 7           & 6700--6928 \\
NGC\,2243                    & 4.365 & 11.00          & $-0.44 \pm 0.05$ & $2.94 \pm 0.10$       & 7           & 6064--6314 (post turn-off) \\
\hline
\end{tabular}
\end{footnotesize}
\tablefoot{
Ages were taken from \citet{2020A&A...640A...1C}. Galactocentric distances were 
computed with the{\fontfamily{cmtt} \selectfont astropy} package (see 
Sect.~\ref{sec:kindyn}). Average [Fe/H] from UVES members in GES iDR6 are 
reported for clusters older than 100 Myr, while for younger objects median (and 
more uncertain) [Fe/H] are given in brackets. The error bars associated with 
the average maximum \element[ ][7]{Li} abundances represent the standard 
deviation of the stars used to compute the mean. 
\tablefoottext{a}{Star forming region. Age and galactocentric distance from 
\citet{2017A&A...601A..70S}.} 
\tablefoottext{b}{ASCC\,16 in \citet{2020A&A...640A...1C}.}
}
\end{table*}

   We estimated the age for each star of our working sample using the aussieq2 
   tool\footnote{\url{https://github.com/spinastro/aussieq2}} that is an 
   extension of the qoyllur-quipu ($q^2$) Python package 
   \citep{2014A&A...572A..48R}. It calculates stellar ages (and masses) by 
   isochrone fitting, starting from the stellar parameters ($T_{\mathrm{eff}}$, 
   $\log g$, and $V$ magnitude) and a grid of isochrones, also taking into 
   account the uncertainties on the stellar parameters. The difference between 
   the observed parameters and the corresponding values in the model grid is 
   used as a weight to calculate the probability distribution function; the 
   most probable age, that is, the peak of the probability distribution, is 
   obtained through a maximum likelihood calculation \citep[see 
     also][]{2020A&A...639A.127C}. For this work, we adopted the Yale-Potsdam 
   Stellar Isochrones \citep[YaPSI;][]{2017ApJ...838..161S} and took the 
   $\alpha$-enhancement effects on the model atmospheres into account, 
   following the procedure outlined in \citet[][their 
     Sect.~2.3]{2020A&A...639A.127C}.

   The recommended GES iDR6 stellar parameters (from spectroscopy), lithium 
   abundances and 
   [$\alpha$/Fe]\footnote{[$\alpha$/Fe]~=~([Ti/Fe]+[Ca/Fe]+[Si/Fe]+[Mg/Fe])/4.} 
   ratios, along with their errors, for the selected stars are reported in 
   Cols.~4 to 13 of Table~\ref{tab:FLDsample}. In general, within the GES 
   collaboration the stellar spectra were processed by different analysis 
   nodes, each adopting the same model atmospheres and line lists though 
   \citep{2015PhyS...90e4010H}, which provided different measurements affected 
   by different biases and random errors \citep{2014A&A...570A.122S}. In order 
   to homogenise the results, {\it Gaia} benchmark stars were used as reference 
   objects to identify the bias function of each node. In GES iDR6, the errors 
   provided by the various nodes were not used -- the errors associated with 
   the recommended values were mainly estimated as internal errors of the 
   method. Lithium abundances, however, were determined only by the Arcetri 
   analysis node (Franciosini et al., in prep.). For this element, the 
   associated uncertainties were derived by combining in quadrature the 
   uncertainties due to the errors on each stellar parameter and on the 
   measured equivalent width \citep[the latter obtained using the formula 
     by][]{1988IAUS..132..345C}. The ages and the galactocentric distances 
   (computed as specified in Sect.~\ref{sec:kindyn}) of objects in our 
   well-controlled sample of Milky Way field stars are listed in the last three 
   columns of Table~\ref{tab:FLDsample}.

   The field star sample was complemented with estimates of the average maximum 
   \element[ ][7]{Li} abundance\footnote{This is the average of the highest 
     \element[ ][7]{Li} abundances that are measured in possibly un-depleted 
     cluster members.} in selected OCs. Ages and distances of OC members are, 
   in general, much more accurate than for isolated field stars. Moreover, 
   young OCs did not have time to travel significantly away from their 
   birthplaces. Therefore, their chemical composition is representative of the 
   history of chemical enrichment of their environs. The GES iDR6 has delivered 
   parameters and abundances for members of 87 OCs and star forming regions, 
   including 20 objects for which spectra were retrieved from the ESO archive 
   and 2 clusters used for calibration only \citep[i.e. not overlapping with 
     the science sample; see][]{2017A&A...598A...5P}. For young clusters 
   present in the previous internal data release of the survey (iDR5), GES 
   spectroscopy and {\it Gaia} astrometry were combined to assign membership 
   probabilities \citep[see][]{2020MNRAS.496.4701J}; targets with probability 
   $P >$~0.9 were selected as cluster members. For a few young clusters not 
   present in GES iDR5, the membership was established based only on radial 
   velocity. For intermediate-age and old clusters, the members were selected 
   as in \citet{2021arXiv210504866M}, deriving first the peak and standard 
   deviation of the radial velocity distribution and selecting objects within 
   2\,$\sigma$ of the peak. Then, the average parallax and proper motion with 
   the corresponding standard deviations were calculated for the selected 
   objects and a 2\,$\sigma$ clipping was applied. By using this method, we 
   found an excellent agreement with the results of \citet{2020A&A...640A...1C} 
   for the clusters in common.

\begin{table*}
\caption{Orbital parameters for the selected clusters and field stars, sorted 
  by increasing age.}
\label{tab:ORBparams}
\centering
\begin{footnotesize}
\begin{tabular}{l c c c c c c c}
\hline\hline
Cluster name/GES name & $U$           & $V$           & $W$          & $J_r$            & $J_z$            & $L_r$             & $L_z$            \\
                      & [km s$^{-1}$] & [km s$^{-1}$]  & [km s$^{-1}$] & [kpc km s$^{-1}$] & [kpc km s$^{-1}$] & [kpc km s$^{-1}$] & [kpc km s$^{-1}$] \\
\hline
\multicolumn{8}{c}{Clusters}\\
\hline
Alessi\,43            & $-21.69$      & $212.26$      & $-0.43$      &  $7.871$         & $0.095$          & $10.792$         & $1724.73$  \\
25\,Ori               & $-6.55$       & $224.21$      & $2.94$       &  $1.488$         & $0.316$          & $18.832$         & $1863.66$  \\
Collinder\,197        & $-22.63$      & $214.22$      & $-1.92$      &  $7.770$         & $0.086$          &  $8.823$         & $-1744.27$ \\
NGC\,2232             & $-14.01$      & $215.96$      & $-4.49$      &  $3.047$         & $0.152$          &  $3.651$         & $1785.27$  \\
NGC\,2547             & $-5.20$       & $221.92$      & $-3.94$      &  $0.501$         & $0.147$          &  $7.550$         & $1783.70$  \\
\ldots                & \ldots        & \ldots        & \ldots       & \ldots           & \ldots           & \ldots           & \ldots     \\
\hline
\multicolumn{8}{c}{Field stars}\\
\hline
17530917-2937418      & $-21.46$      & $223.21$      & $-11.88$     &  $4.453$         & $0.785$          & $10.015$         & $1341.21$ \\
17524903-2926366      & $-29.35$      & $237.86$      & $-2.50$      & $12.317$         & $0.118$          &  $9.626$         & $1407.15$ \\
10371153-5838408      & $-34.89$      & $182.58$      & $-8.17$      & $47.240$         & $0.415$          &  $2.812$         & $1340.94$ \\
17530416-2936240      & $20.21$       & $217.83$      & $9.66$       &  $3.166$         & $0.990$          & $19.701$         &  $962.62$ \\
17525280-2931432      & $36.47$       & $238.86$      & $-16.62$     & $19.449$         & $1.417$          &  $8.693$         & $1472.30$ \\
\ldots                &\ldots         & \ldots        & \ldots        & \ldots          & \ldots           & \ldots           & \ldots    \\
\hline
\end{tabular}
\end{footnotesize}
\tablefoot{
Columns 2 to 4 report the galactocentric velocity components; Cols.~5 and 6 
give the radial and vertical actions, respectively; Cols.~7 and 8 report the 
radial and vertical components of the angular momentum. The table is available 
in its entirety at the CDS. A portion is shown here for guidance regarding its 
form and contents.
}
\end{table*}

   \begin{figure}
     \centering
     \includegraphics[width=7cm]{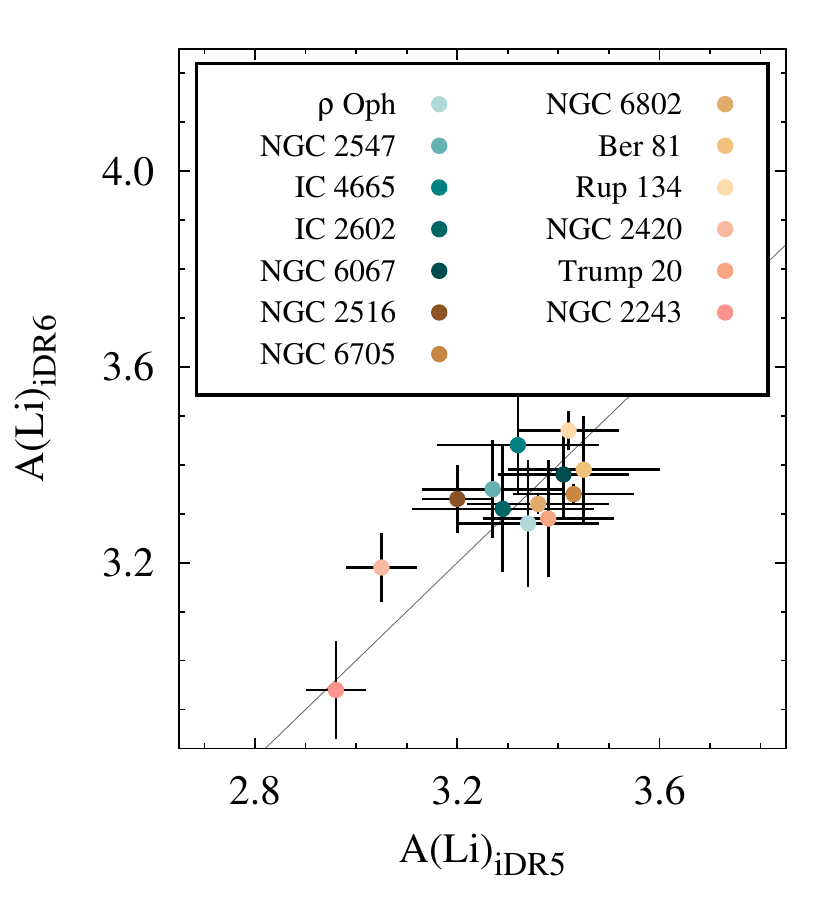}
     \caption{ Average maximum \element[ ][7]{Li} abundances from iDR6 (this 
       work) and iDR5 \citep{2020A&A...640L...1R} for the clusters in common.}
     \label{fig:diffALimax}
   \end{figure}

   Following the criteria outlined by \citet{2020A&A...640L...1R} and briefly 
   recalled below, we restricted our analysis to 26 clusters including one star 
   forming region with age less than 5~Myr and several young OCs (age~$<$ 
   100~Myr), paying attention to select for each cluster only the members that 
   suffered minimal lithium depletion, if any (see Table~\ref{tab:OCsample}). 
   In the first place, we selected very young clusters with \element[ ][7]{Li} 
   detections in at least four to five members that supposedly reflect the 
   original \element[ ][7]{Li} abundance of their parent clouds. Clusters 
   younger than 100~Myr host pre-main-sequence or zero-age main-sequence stars 
   that should not have depleted any \element[ ][7]{Li}. As a matter of fact, 
   though, \element[ ][7]{Li} depletion may show up in cool members of clusters 
   as young as $\sim$5~Myr \citep[see e.g.][and references 
     therein]{2016A&A...590A..78B,2021MNRAS.500.1158J}. The average maximum 
   \element[ ][7]{Li} abundance for the young clusters considered in this work 
   was computed selecting only the stars that trace the upper envelope of the 
   $A$(Li) versus $T_{\mathrm{eff}}$ distributions, for GES clusters sampled well 
   enough, which maximises our chances to recover the initial, un-depleted 
   value. We added to this sample clusters in which stars on the blue (warm) 
   side of the `lithium dip' \citep{1986ApJ...302L..49B} were observed, either 
   on the main sequence or after the turn-off, before surface dilution in 
   connection with the first dredge-up starts. The only exception is NGC\,2243 
   that, because of its older age ($\simeq 4.4$~Gyr, all the other clusters 
   being younger than 2~Gyr), is likely to display only a lower limit to the 
   actual original value. With this caveat in mind, we retained this cluster to 
   increase the statistics in the outer disc. The average maximum 
   \element[ ][7]{Li} abundance, $A$(Li)$_\mathrm{max}$, along with the 
   associated dispersion, the age, galactocentric distance and metallicity 
   adopted for each cluster are listed in Table~\ref{tab:OCsample}. The number 
   of stars used to calculate $A$(Li)$_\mathrm{max}$ and their $T_\mathrm{eff}$ 
   range are also given in Cols.~6 and 7. Notwithstanding the relative paucity 
   of objects that passed our selection criteria (30 per cent of all OCs in GES 
   iDR6), we can still probe a wide galactocentric distance baseline, $5 < 
   R_\mathrm{GC}/\mathrm{kpc} < 15$, which makes it possible to do a meaningful 
   test of the predictions of the GCE model regarding the existence and extent 
   of any \element[ ][7]{Li} gradient across the Milky Way disc.

   For most clusters in common, the average maximum \element[ ][7]{Li} 
   abundances displayed in Table~\ref{tab:OCsample} differ slightly (see 
   Fig.~\ref{fig:diffALimax}) from the corresponding values reported in Table~1 
   of \citet{2020A&A...640L...1R}. This is mainly due to the improved abundance 
   measurements in iDR6 with respect to the previous internal data release of 
   the GES on which the work of \citeauthor{2020A&A...640L...1R} is based, 
   though in some cases the number of member stars with \element[ ][7]{Li} 
   determinations used to compute the mean has been reduced. It is further 
   noted that some very young clusters studied by \citet{2020A&A...640L...1R} 
   -- namely, NGC\,6530, Trumpler\,14, Chamaeleon~I, and NGC~2264 -- show a 
   large scatter in their revised \element[ ][7]{Li} abundances, possibly due 
   to stellar activity or accretion, while a few interesting objects included 
   in iDR6 (e.g. NGC\,6649 and NGC\,6281) are still missing a clean membership 
   analysis. These clusters are not part of the present study.

   \subsection{Kinematic and dynamical properties of OCs and field stars}
   \label{sec:kindyn}

   We set the distance of the Sun from the Galactic centre to $R_0 =$ 8~kpc,
   its height above the plane to $z_0 =$ 0.025~kpc \citep{2008ApJ...673..864J,
     2012ApJ...759..131B}, and used the astrometric data from {\it Gaia} to 
   transform the celestial coordinates of stars and OCs in our sample in 
   galactocentric radius and height above the Galactic plane 
   using{\fontfamily{cmtt} \selectfont 
     astropy}\footnote{\url{https://www.astropy.org/}} 
   \citep{2013A&A...558A..33A}. The radial velocities from GES iDR6 and 
   parallaxes and proper motions from {\it Gaia} were used to obtain the 
   orbital parameters and actions of clusters and field stars in our working 
   sample with the{\fontfamily{cmtt} \selectfont galpy} 
   package\footnote{\url{https://github.com/jobovy/galpy}} 
   \citep{2015ApJS..216...29B}. For the Milky Way's gravitational potential, we 
   assumed the code built-in model{\fontfamily{cmtt} \selectfont 
     MWpotential2014}. The local standard of rest (LSR) velocity was set to 
   $V_{\mathrm{LSR}} =$ 220~km~s$^{-1}$ \citep{2012ApJ...759..131B} and we 
   assumed ($U$, $V$, $W$)$_\sun$~=~(11.1, 12.24, 7.25) km~s$^{-1}$ for the 
   velocity of the Sun relative to the LSR \citep{2010MNRAS.403.1829S}. The 
   orbital parameters of our sample clusters and stars that we exploit in 
   Sect.~\ref{sec:dyn} are listed in Table~\ref{tab:ORBparams}.

   \section{Chemical evolution model}
   \label{sec:model}

   This section provides an overview of the adopted GCE model. More information 
   can be found in the original papers.

   \subsection{General features}

   In order to study the evolution of lithium in the Milky Way, we adopted the 
   `parallel GCE model' presented by \citet{2017MNRAS.472.3637G,
     2018MNRAS.481.2570G}. In the parallel model, the thick- and thin-disc 
   components of our Galaxy are formed on different timescales out of two 
   distinct episodes of infall of primordial gas and evolved independently from 
   each other \citep[see \citealt{1992ApJ...387..138F} and][for former 
     approaches to the parallel galaxy formation scenario]{2009IAUS..254..191C}.

   The contemporary evolution of the two components causes their stellar 
   populations to overlap in a wide range of metallicities, while the less 
   (more) efficient star formation assumed for the thin (thick) disc leads to 
   distinct sequences of lower (higher) [$\alpha$/Fe] ratios in the 
   [$\alpha$/Fe] versus [Fe/H] plane, as observed \citep[see][and references 
     therein]{2017MNRAS.472.3637G}. The interested reader is referred to 
   \citet{2017MNRAS.472.3637G,2018MNRAS.481.2570G} for a detailed description 
   of the main assumptions and underlying equations of the model. In the 
   following, we focus on the adopted nucleosynthesis prescriptions, with 
   special emphasis on those for \element[ ][7]{Li}. 

   \subsection{Nucleosynthesis prescriptions}
   \label{sec:nuc}

   The nucleosynthesis prescriptions are the same as model MWG-11 of 
   \citet{2019MNRAS.490.2838R}. For low- and intermediate-mass stars with 
   initial masses in the range $1 \le m/{\mathrm M}_\sun < 9$, we adopted the 
   yields by \citet{2013MNRAS.431.3642V,2020A&A...641A.103V} that include a 
   proper treatment of the super-AGB phase for the most massive AGB stars. For 
   massive stars with $13 \le m/{\mathrm M}_\sun \le 100$, we adopted the 
   yields by \citet{2018ApJS..237...13L}. In particular, for [Fe/H]~$\le -1$ 
   dex we used their `set R' (see the original paper for details) and assumed 
   that all massive stars are fast rotators, whilst for [Fe/H]~$> -1$ dex we 
   used the yields from the non-rotating stellar models. With these choices we 
   are able to fit the abundance measurements of several chemical elements in 
   the solar neighbourhood as well as across the whole Milky Way disc 
   \citep[see][Baratella et al., in prep.; Grisoni et al., in 
     prep.]{2019MNRAS.490.2838R,2020A&A...639A..37R,2020MNRAS.498.1252G}. Thus, 
   we believe that the adopted yield set provides a robust backbone for general 
   GCE studies. Since the $\nu$-process is not included in the computations of 
   \citet{2018ApJS..237...13L}, to take account of the effects of 
   neutrino-induced nucleosynthesis on the evolution of \element[ ][7]{Li} in 
   the Galaxy we resorted to the \element[ ][7]{Li} yields by 
   \citet{1995ApJS..101..181W}\footnote{Although other authors 
     \citep[e.g.][]{2008ApJ...686..448Y,2018ApJ...865..143S,2019ApJ...876..151S,
       2019ApJ...872..164K} provide up-to-date estimates of the $\nu$-process 
     \element[ ][7]{Li} yield using cutting-edge SN explosion and SN neutrino 
     models, as well as new cross-sections for the relevant reactions, the work 
     by \citet{1995ApJS..101..181W} remains the only one to offer a grid of 
     yields that span a range of stellar masses and metallicities adequate for 
     use in GCE studies.}. Following arguments by \citet{1997ApJ...488..338D}, 
   we lowered the original yields by a factor of two \citep[see discussion 
     in][]{2001A&A...374..646R}.

   The stellar yields for type Ia SNe (exploding WDs in binary systems) were 
   taken from \citet[][their model W7]{1999ApJS..125..439I}. Type Ia SNe are 
   responsible for the bulk of Fe production at late times and dominate the 
   decrease in the [$\alpha$/Fe] ratio at high metallicities, but do not 
   produce any \element[ ][7]{Li}. Classical novae, instead, are possibly 
   important \element[ ][7]{Li} factories. The nova outburst rate was 
   implemented in the parallel GCE code following the scheme outlined in 
   \citet[][see also \citealt{1991A&A...248...62D} and 
     \citealt{1995A&A...303..460M}]{1999A&A...352..117R}. The fiducial model 
   adopted in this work assumes that nearly 2 per cent of all stars with 
   initial mass in the range 1--8~M$_\sun$ enter the formation of nova systems. 
   Unless otherwise stated, the average \element[ ][7]{Li} production per nova 
   outburst is set to $M^\mathrm{nova}_\mathrm{Li,\,burst} = 2.55 \times 
   10^{-10}$~M$_\sun$, consistent with the direct estimate of the 
   \element[ ][7]{Li} yield from observations of nova V1369\,Cen by 
   \citet{2015ApJ...808L..14I} and with the results from some recent 
   hydrodynamical simulations \citep{2020ApJ...895...70S}. When assuming that 
   each nova system suffers 10$^4$ outbursts during its lifetime 
   \citep{1978MNRAS.183..515B}, the above number translates into a total 
   ejected mass of \element[ ][7]{Li} per nova of 
   $M^\mathrm{nova}_\mathrm{Li,\,tot} \simeq$ 2.5~$\times 10^{-6}$~M$_\sun$. We 
   explore the effects of deviations from the main assumptions of this 
   fiducial model in Sect.~\ref{sec:change}.

   Cosmic-ray bombardment of the ISM also creates \element[ ][7]{Li}. The 
   contribution of GCR spallation was taken into account by incorporating the 
   absolute yields by \citet{1998ApJ...499..735L}. We took the yields 
   corresponding to the lower-bound spectrum in their Table 1. Alternatively, 
   one could use empirical yields inferred from \element[ ][9]{Be} observations 
   in stars to find similar results \citep[see][and references 
     therein]{2019MNRAS.489.3539G}.

   \section{Results}
   \label{sec:results}

   \begin{figure*}[ht]
     \centering
     \includegraphics[width=14cm]{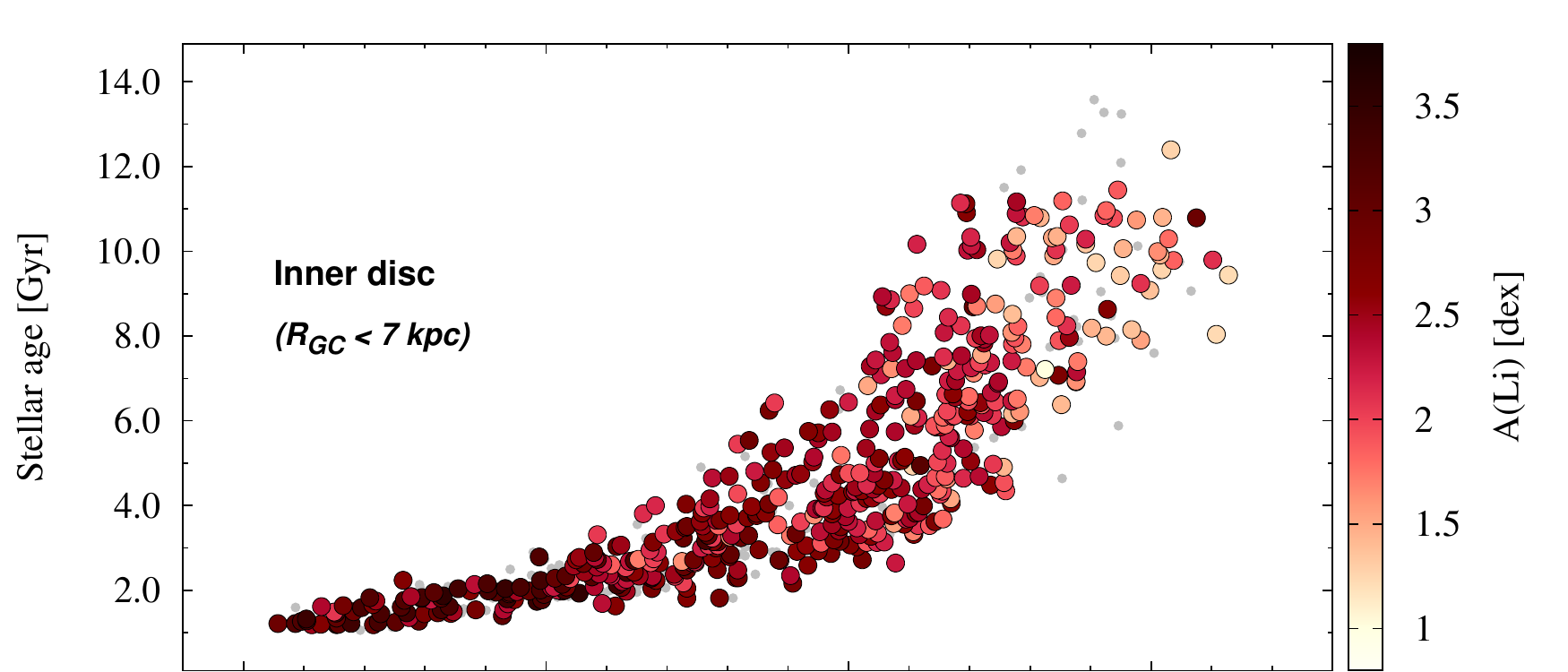}
     \includegraphics[width=14cm]{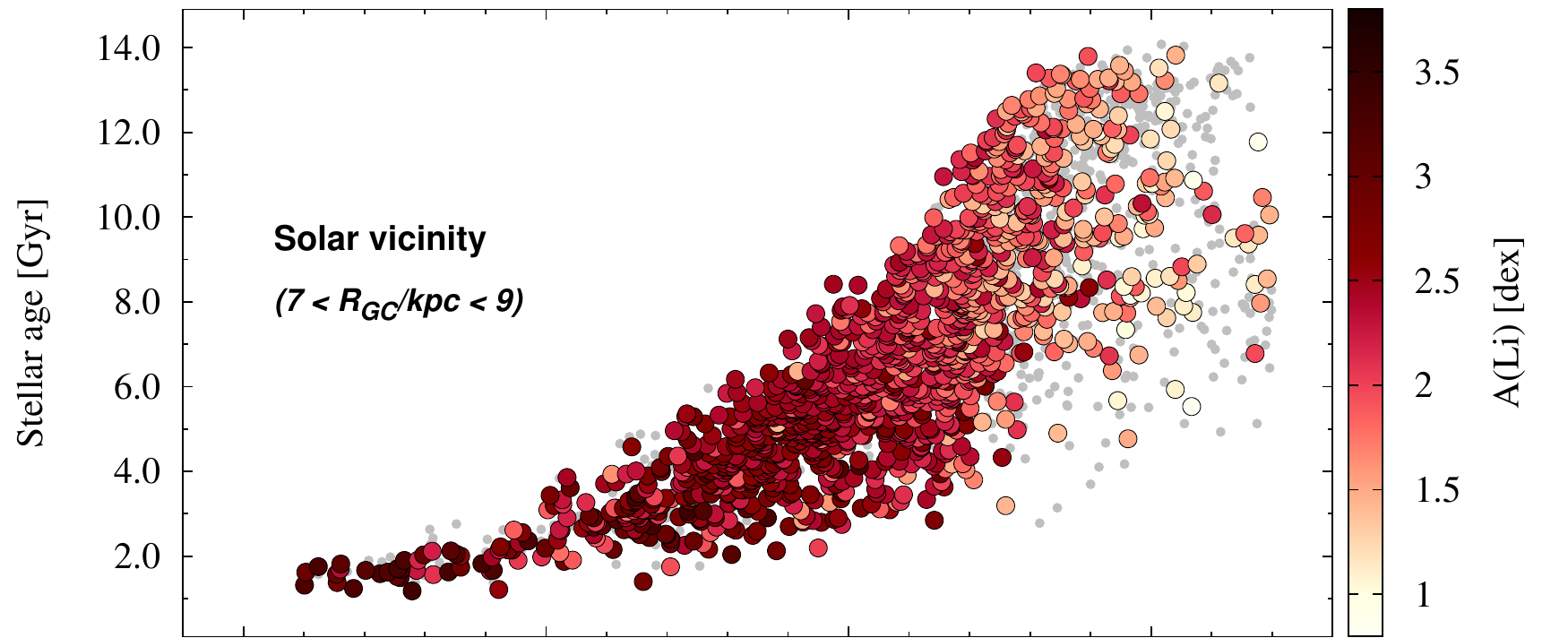}
     \includegraphics[width=14cm]{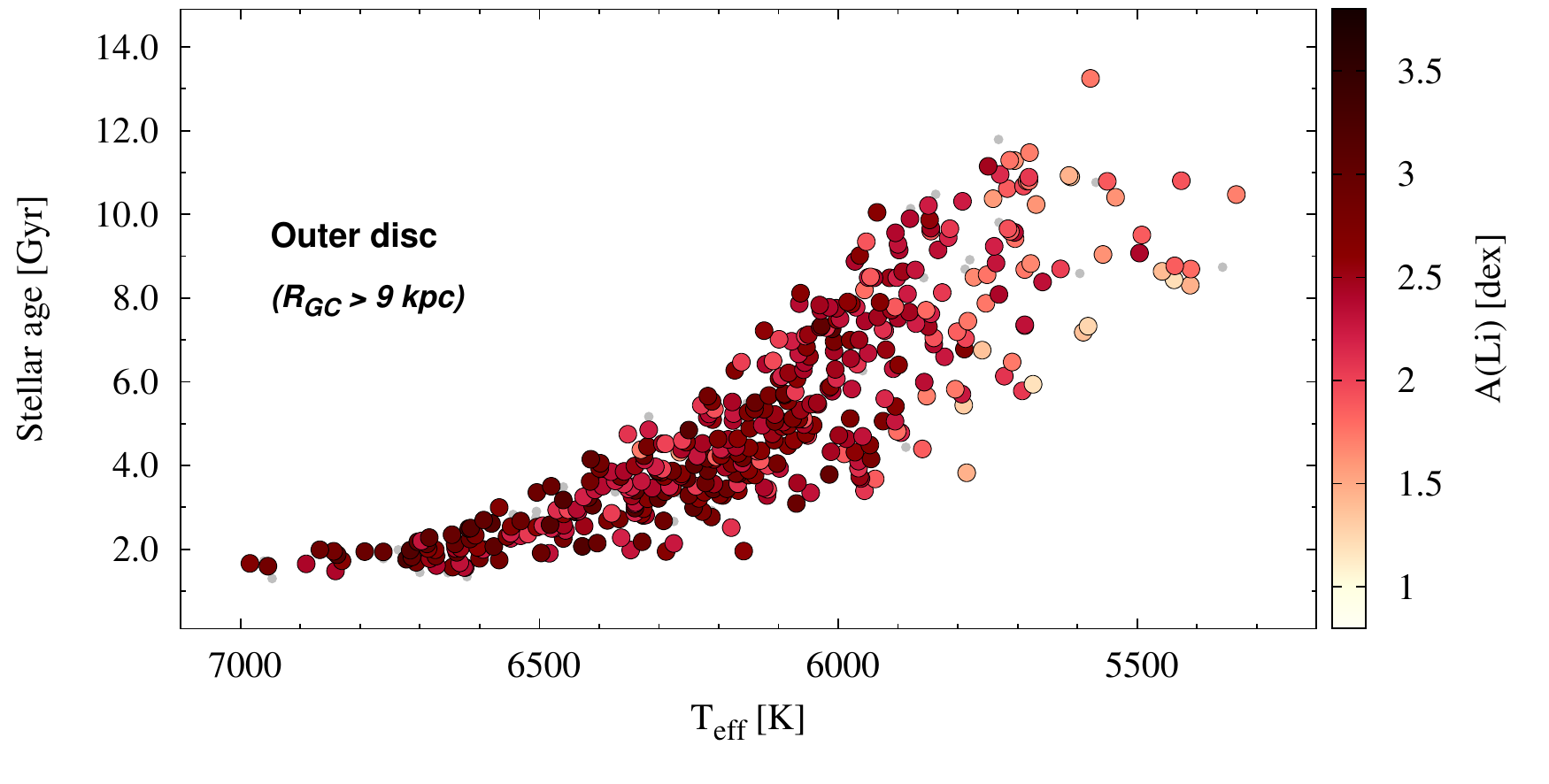}
     \caption{ Effective temperature versus stellar age diagrams for the GES 
       iDR6 sample of field stars discussed in this paper, divided into inner 
       disc (\emph{top panel,} 619 stars), solar vicinity (\emph{middle panel,} 
       2129 stars), and outer disc sub-samples (\emph{bottom panel,} 462 
       stars). Stellar ages were determined from isochrone fitting using the 
       aussieq2 code and YaPSI stellar models \citep{2017ApJ...838..161S}. 
       Stars with lithium measurements are colour-coded according to their 
       \element[ ][7]{Li} abundance. Grey dots represent upper limits.}
     \label{fig:AgeTeff}
   \end{figure*}

   \begin{figure*}[ht]
     \centering
     \includegraphics[width=14cm]{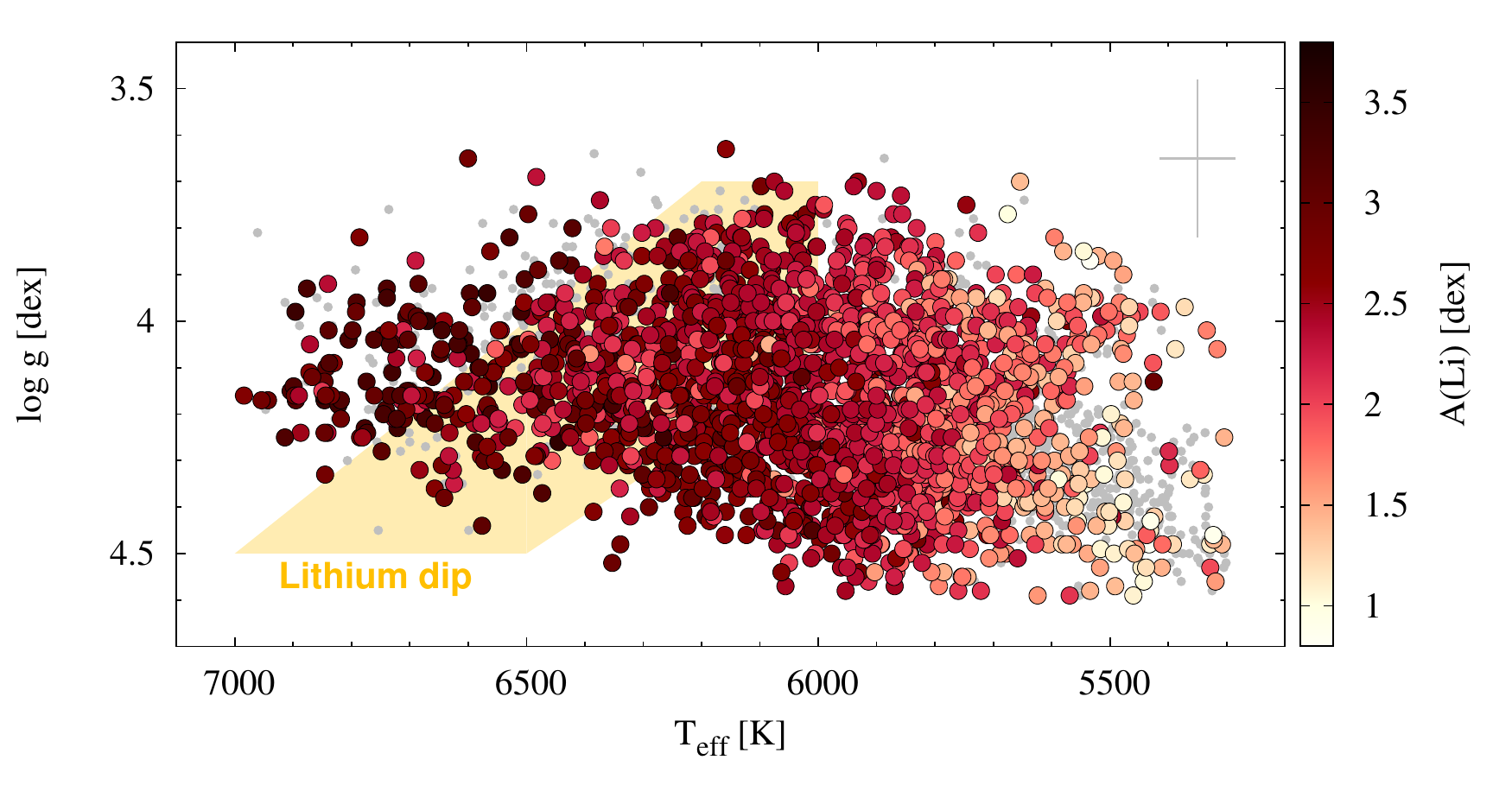}
     \caption{ Effective temperature versus gravity diagram for the full GES 
       iDR6 sample of field stars discussed in this paper. Stars with lithium 
       measurements are colour-coded according to their \element[ ][7]{Li} 
       abundance. Grey dots represent upper limits. The approximate location of 
       the \element[ ][7]{Li} dip region \citep[see 
         also][]{2020MNRAS.497L..30G} is highlighted (yellow area).}
     \label{fig:LoggTeff}
   \end{figure*}

   In this section we exploit the GES iDR6 dataset presented in 
   Sect.~\ref{sec:data} and the GCE model introduced in Sect.~\ref{sec:model} 
   to study the evolution of lithium in the Milky Way. We first present the 
   trends of $A$(Li) with stellar parameters for our sample of field stars, 
   divided into inner disc, solar neighbourhood, and outer disc inhabitants 
   according to their current positions along the disc. In Sect.~\ref{sec:sn}, 
   we focus on \element[ ][7]{Li} enrichment in the solar vicinity. Solar 
   neighbourhood stars in our sample are separated into four distinct 
   populations via chemical selection criteria and their properties are 
   compared to the predictions of the parallel GCE model. We then extend our 
   modelling to the whole Milky Way disc to investigate the Galactic 
   \element[ ][7]{Li} gradient. Lastly, we use the Toomre diagram and a diagram 
   of angular momentum, $L_z$, and square root of the radial action, 
   $\sqrt{J_r}$ \citep[see][]{2020MNRAS.497..109F}, to assess the membership of 
   the stars to the thick or thin discs on dynamical bases and to pinpoint 
   possibly accreted stars. The lithium content of a few candidate accreted 
   stars is discussed in comparison to the results of models that trace the 
   evolution of \element[ ][7]{Li} in local dwarf galaxies 
   \citep{2021arXiv210411504M}.

   \subsection{Tracing the evolution of lithium in the Milky Way}
   \label{sec:trace}

   Figures~\ref{fig:AgeTeff} and \ref{fig:LoggTeff} show, respectively, the 
   $T_\mathrm{eff}$ versus age and Kiel diagrams for our selection of field 
   stars with \element[ ][7]{Li} determinations from high-quality spectra 
   (S/N~$\ge$ 50) in GES iDR6 catalogue. In agreement with previous studies 
   \citep[see e.g.][for recent work]{2018A&A...615A.151B}, we find that: (i) 
   due to the weakness of the \element[ ][7]{Li} line, several high-temperature 
   stars have only upper limits on their \element[ ][7]{Li} abundance; (ii) 
   \element[ ][7]{Li} measurements in stars on the lower main sequence witness 
   efficient lithium destruction in the deeper convective envelopes of the 
   cooler stars; (iii) the highest \element[ ][7]{Li} abundances, $A$(Li)~$> 
   3.0$~dex, are found in the upper main sequence and around the turn-off; (iv) 
   in the lithium dip region, the stars span the full range of 
   \element[ ][7]{Li} abundances, from the Spite plateau value to very close to 
   the meteoritic one, with a few objects characterised by anomalously low 
   lithium abundances; (v) the hotter the stars, the less we can go back in 
   time. Referring to the last point, we note that our field star sample does 
   not contain any objects younger than 1~Gyr, due to GES selection criteria 
   \citep[see also][]{2018MNRAS.473..185T}. We further remark that the ages of 
   the cool stars are very uncertain, because of the fact that the isochrones 
   tend to crowd together. However, this does not impact the present study, 
   since we are mostly interested in the stars that potentially preserve their 
   original \element[ ][7]{Li} and the cool stars are known to efficiently 
   destroy it.

   \begin{figure*}
     \centering
     \includegraphics[width=14cm]{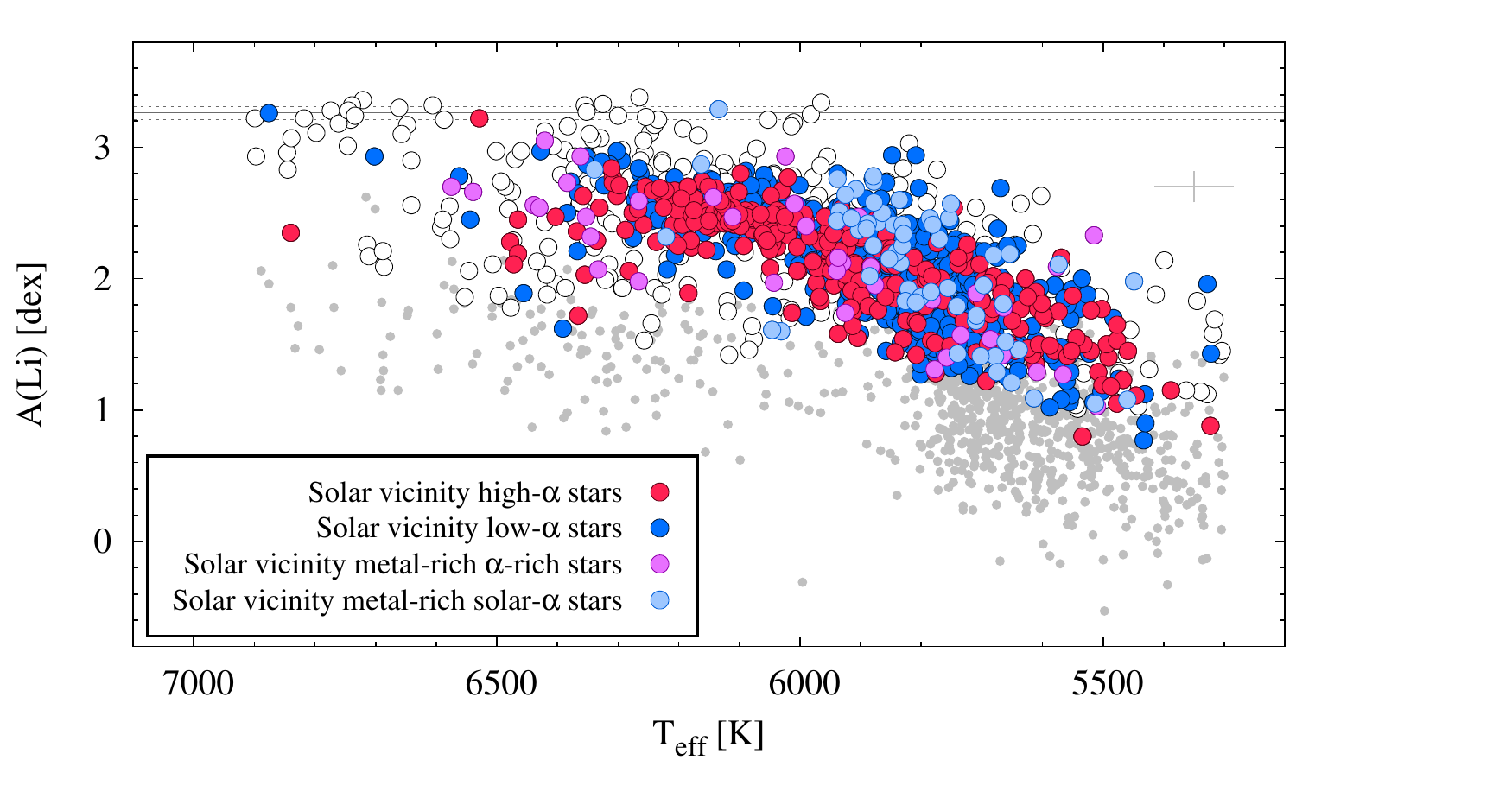}
     \caption{ Lithium abundance as a function of effective temperature for our 
       sample of GES iDR6 field stars restricted to targets found in the solar 
       neighbourhood ($7 < R_{\mathrm{GC}}$/kpc~$< 9$). Both \element[ ][7]{Li} 
       measurements (circles) and upper limits (grey dots) are reported. 
       Whenever possible, high-$\alpha$ (red circles), low-$\alpha$ (blue), 
       metal-rich $\alpha$-rich (magenta), and super metal-rich solar-$\alpha$ 
       stars (light blue) were identified according to their position in the 
       [$\alpha$/Fe]--[Fe/H] diagram, adopting the same criteria as 
       \citet[][see also \citealt{2014A&A...567A...5R}]{2016A&A...595A..18G,
         2019A&A...623A..99G}. Empty circles represent stars without chemical 
       identification. The grey lines show the \element[ ][7]{Li} content in 
       meteorites \citep[$3.26 \pm 0.05$,][]{2009LanB...4B..712L}.}
     \label{fig:ALiTeff}
   \end{figure*}

   After careful scrutiny of Figs.~\ref{fig:AgeTeff} and \ref{fig:LoggTeff}, we 
   are led to suggest that, among the field stars, only those with 
   $T_\mathrm{eff} \ge$ 6800--6900~K can be safely used as tracers of 
   \element[ ][7]{Li} evolution. Stars with $6500 < T_\mathrm{eff}$/K~$< 6800$ 
   could still be employed, but one should consider their location in a Kiel 
   diagram very carefully. Without a severe temperature cut, the commonly 
   adopted approach of binning the data in metallicity to compute mean values 
   of $A$(Li) for the stars with the highest \element[ ][7]{Li} abundances in 
   each metallicity bin will lead to a spurious $A$(Li) versus [Fe/H] trend 
   either way, with the computed values unavoidably reflecting stellar 
   depletion to some extent \citep{2018AJ....155..138A}. Our suggestion is 
   consistent with the findings of \citet{2020MNRAS.497L..30G}. These authors 
   separate nearly 63\,000 late-type stars with \element[ ][7]{Li} abundances 
   obtained in the framework of the Galactic Archeology with HERMES (GALAH) 
   survey into two groups -- one on the warm side, the other on the cool side 
   of the lithium dip -- and analyse the behaviour of each group in the 
   $A$(Li)--[Fe/H] plane. They suggest that the warm stars trace effectively 
   the evolution of \element[ ][7]{Li} in the Galaxy, while the cool stars have 
   burnt lithium to a large degree \citep[see also][]{2021A&A...649L..10C}.

   \begin{figure*}[ht]
     \centering
     \includegraphics[width=14cm]{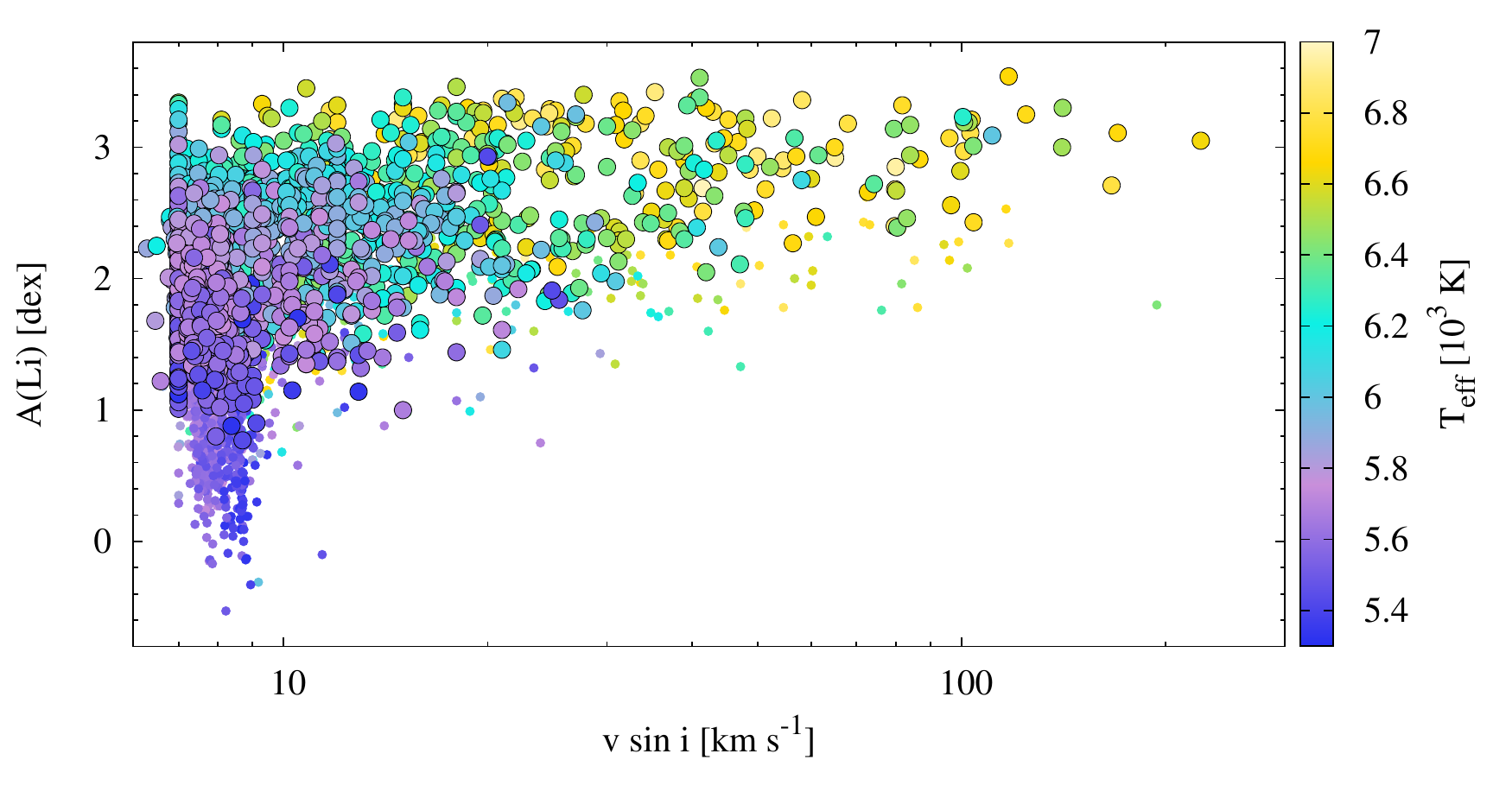}
          \caption{ Lithium abundance as a function of rotational velocity for 
            the full field GES iDR6 stellar sample selected for this paper 
            (3210 objects). The stars are colour-coded according to their 
            effective temperature. Both \element[ ][7]{Li} measurements 
            (circles) and upper limits (dots) are shown. Rotational velocities 
            are measured from FLAMES-UVES spectra; typical errors on 
            $\vel\sin$\,i are $\le 5$ km~s$^{-1}$.}
     \label{fig:ALivsini}
   \end{figure*}

   In Fig.~\ref{fig:ALiTeff} we display the run of $A$(Li) with $T_\mathrm{eff}$ 
   for stars currently found in the solar neighbourhood. Our sample contains 
   1293 solar neighbourhood stars with \element[ ][7]{Li} measurements (for 836 
   targets we could only give upper limits) that belong to chemically distinct 
   populations. We adopted the criteria presented in 
   \citet{2014A&A...567A...5R} and applied later to lithium studies in 
   \citet{2016A&A...595A..18G,2019A&A...623A..99G} and identified 311 
   high-$\alpha$ stars, 524 low-$\alpha$ stars, a metal-rich $\alpha$-rich 
   population \citep[also called mr$\alpha$r,][]{2011A&A...535L..11A} 
   consisting of 41 stars, and a super metal-rich solar-$\alpha$ population (61 
   objects)\footnote{For the sake of brevity, we name stars displaying high 
     [$\alpha$/Fe] ratios, high-$\alpha$ stars, those displaying low 
     [$\alpha$/Fe] ratios, low-$\alpha$ stars, etcetera.}. We did not classify 
   356 stars that lack simultaneous measurements of the elements used to define 
   the [$\alpha$/Fe] ratios in this work (Mg, Si, Ca, Ti; empty circles in 
   Fig.~\ref{fig:ALiTeff}). All the sub-populations are characterised by 
   similar $A$(Li) spreads at a given temperature (when considering only the 
   temperature range in which all populations are well sampled). In the range 
   $5500 < T_{\mathrm{eff}}$/K~$< 6300$, the upper envelope of the 
   \element[ ][7]{Li} abundances traced by the high-$\alpha$ stars lies 
   systematically below that traced by the other components. If the 
   high-$\alpha$ stars were all born in the thick-disc component, this could 
   indicate that the thick disc attained a lower level of \element[ ][7]{Li} 
   enrichment \citep[see][]{2016A&A...595A..18G,2019MNRAS.489.3539G}. This 
   conclusion is reinforced by an inspection of the $A$(Li) versus [Fe/H] 
   diagram (see next section). We further note that there is a temperature 
   threshold, $T_{\mathrm{eff}} \simeq 6000$~K, above which the stars with the 
   highest \element[ ][7]{Li} abundances in our solar vicinity sample share the 
   same (meteoritic) abundance; below such a threshold the abundance of lithium 
   decreases steadily with decreasing temperature, as expected.

   We also looked for possible correlations between \element[ ][7]{Li} 
   abundance and rotation in our sample of field stars. 
   Figure~\ref{fig:ALivsini} shows the $A$(Li)--$\vel\sin$\,i plane, where 
   $\vel\sin$\,i is the projected equatorial rotational velocity, for all the 
   3210 stars with high-quality \element[ ][7]{Li} abundances in our sample 
   colour-coded in accordance with their effective temperature. Unsurprisingly, 
   the hottest (more massive) stars in the sample have the highest rotational 
   velocities. For $\vel\sin$\,i~$> 30$~km~s$^{-1}$, there is a spread of about 
   1~dex in \element[ ][7]{Li} abundances. The stars in this region of the 
   diagram are all F-type stars; their atmospheres are expected to be mostly in 
   radiative equilibrium and to show minimal convection effects. We interpret 
   the observed spread as due to rotational mixing; the few F-type stars with 
   low $\vel\sin$\,i could have a low inclination angle. For $\vel\sin$\,i~$\le 
   20$~km s$^{-1}$ a much larger spread in \element[ ][7]{Li} abundance is 
   seen, driven by convection. The few stars with $A$(Li) abundances in excess 
   of 3.4~dex are all relatively young (ages~$\sim$~1--2 Gyr) and located in 
   the inner disc, where the GCE model predicts a current ISM 
   \element[ ][7]{Li} abundance higher than in the solar neighbourhood. By 
   sure, such stars reflect an evolutionary path different from that followed 
   by solar neighbourhood stars (see Sect.~\ref{sec:grad}); however, chemical 
   separation may have changed their original composition. In particular, 
   atomic diffusion could lower the surface \element[ ][7]{Li} abundance 
   inherited at birth \citep{2021A&A...649L..10C}. We note that the star with 
   the highest \element[ ][7]{Li} abundance in our sample, $A$(Li)~$= 3.54 \pm 
   0.06$~dex, and with [Fe/H]~$= 0.05 \pm 0.06$ is relatively hot 
   ($T_{\mathrm{eff}} = 6670 \pm 66$~K) and has $\log g = 4.0 \pm 0.17$~dex, that 
   is, it falls outside the Li-dip region according to Fig.~\ref{fig:LoggTeff}. 
   It also has a good S/N of 113 and $\vel\sin$\,i~$= 117.4 \pm 1.1$, which 
   enhances the chances that we are seeing its original lithium content 
   unaltered by mixing processes. In fact, as discussed by 
   \citet{2021A&A...649L..10C}, high rotation rates would counteract atomic 
   diffusion. The information on stellar rotation is thus very useful in that 
   it provides hints on the possible role played by internal \element[ ][7]{Li} 
   depletion processes in warm, metal-rich main-sequence stars.

   \subsection{Evolution of lithium in the solar vicinity}
   \label{sec:sn}

   \begin{figure*}[ht]
     \centering
     \includegraphics[width=6.9cm]{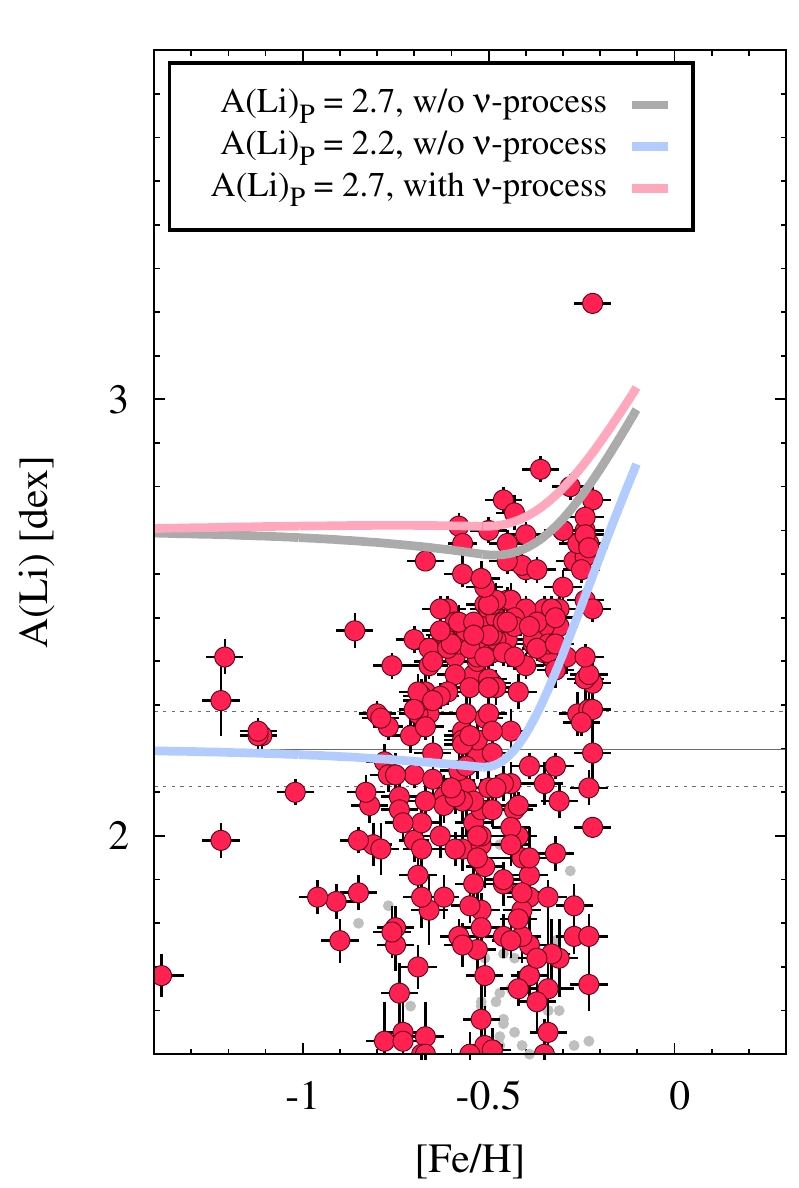}
     \includegraphics[width=5.58cm]{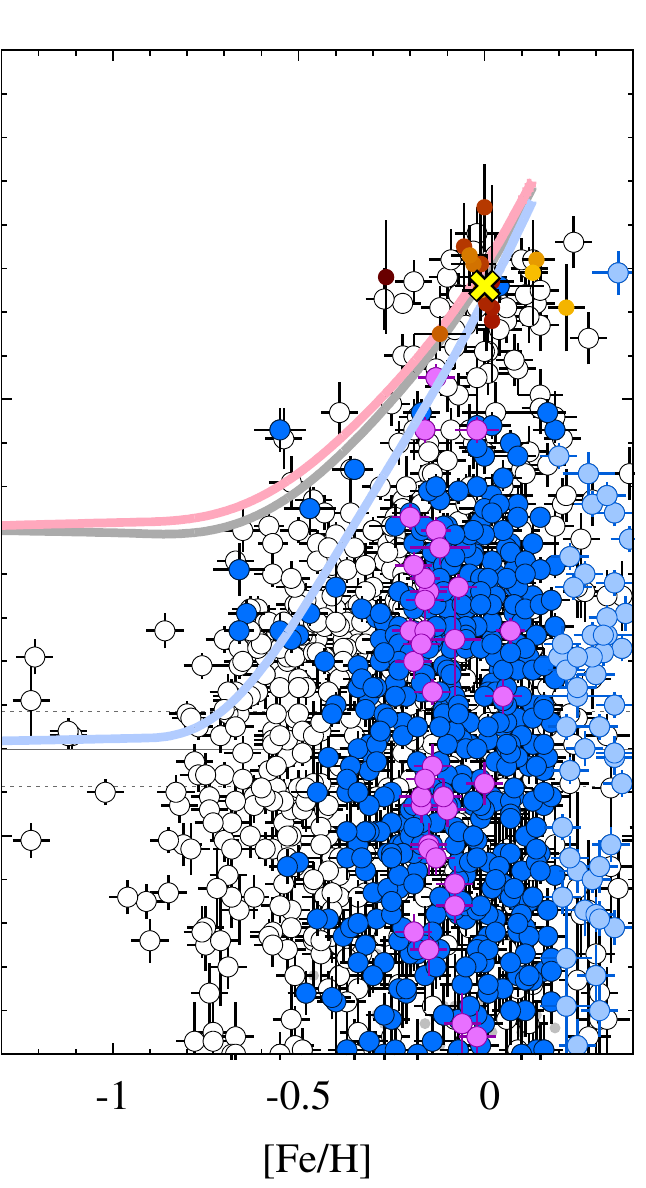}
     \includegraphics[width=5.58cm]{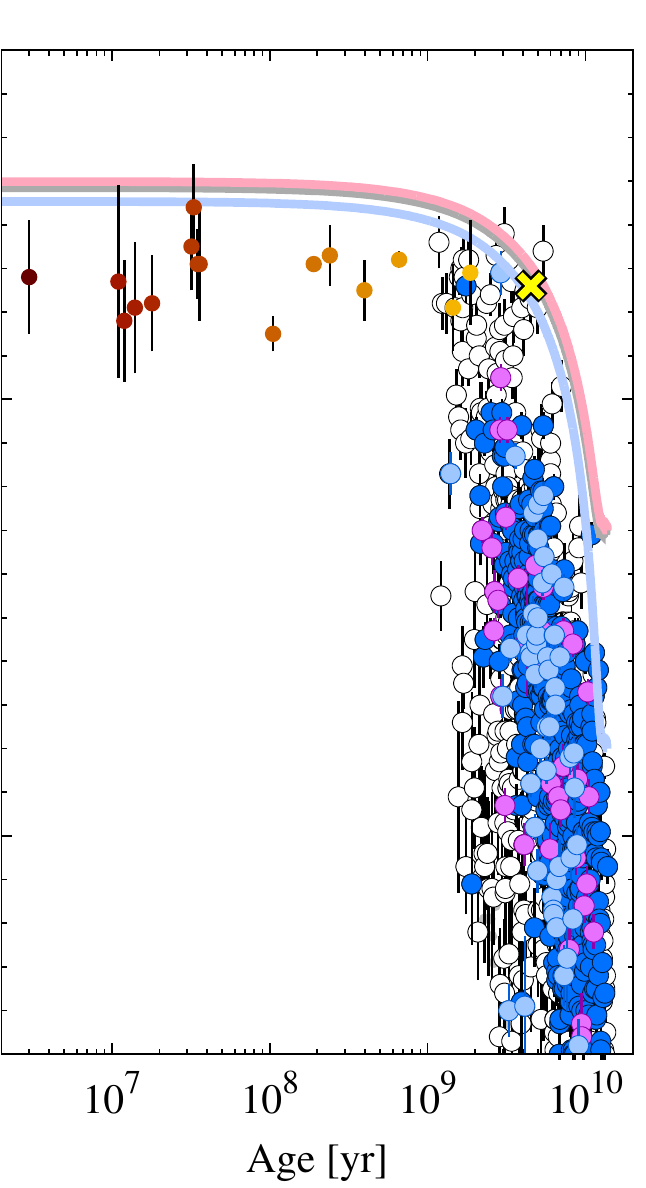}
     \caption{ Behaviour of $A$(Li) as a function of [Fe/H] \emph{(left and 
         middle panels)} and age \emph{(right panel)} in the solar 
       neighbourhood, for the thick- \emph{(left)} and thin-disc components 
       \emph{(middle and right panels)}. Lithium measurements for high- and 
       low-$\alpha$ field stars in GES iDR6 are shown (red and blue circles, 
       respectively), together with upper limits (grey dots). Cluster data are 
       shown with brown shades for the younger clusters and yellow tones for 
       the older ones. Only clusters and stars with $7 < R_\mathrm{GC}$/kpc $< 
       9$ are considered for the comparison with the solar neighbourhood 
       models. The meteoritic abundance from \citet{2009LanB...4B..712L} is 
       indicated with the multiplication sign. The predictions of the parallel 
       GCE model are highlighted with different colours depending on the 
       assumptions made about \element[ ][7]{Li} synthesis and cosmological 
       lithium abundance. We also show data for metal-rich $\alpha$-rich stars 
       (magenta circles) and super metal-rich solar-$\alpha$ stars (light blue 
       circles) found in the solar neighbourhood. These are not to be compared 
       with the predictions of solar neighbourhood models and are used only for 
       discussion purposes (see text). Empty circles are stars that lack an 
       [$\alpha$/Fe] determination in GES iDR6; we assign most of them to the 
       thin-disc component, based on their kinematics (see 
       Sect.~\ref{sec:dyn}).}
     \label{fig:SN}
   \end{figure*}

   In this section, we focus on solar vicinity targets (i.e. clusters and stars 
   with galactocentric distances $7 < R_\mathrm{GC}/\mathrm{kpc} < 9$). In 
   Fig.~\ref{fig:SN} we compare the predictions of the parallel GCE model to 
   \element[ ][7]{Li} abundances of field stars and to average maximum 
   \element[ ][7]{Li} abundances of OCs selected from GES iDR6 (see 
   Sect.~\ref{sec:data}). We assumed either the Spite plateau value, 
   $A$(Li)$_\mathrm{P}^{\mathrm{obs}} \simeq$ 2.2~dex, or the BBN-predicted value, 
   $A$(Li)$_\mathrm{P}^{\mathrm{th}} \simeq$ 2.7~dex, as the initial 
   \element[ ][7]{Li} abundance in the GCE model. The contributions to 
   \element[ ][7]{Li} synthesis from RGB and AGB stars\footnote{Possible 
     production or destruction of \element[ ][7]{Li} due to extra mixing 
     processes, such as thermohaline mixing in low-mass, bright RGB stars and 
     stellar rotation \citep{2020A&A...633A..34C}, are not included in the 
     adopted stellar models and yields \citep{2013MNRAS.431.3642V,
       2020A&A...641A.103V}.}, nova systems, and GCRs were included in all 
   models, following the prescriptions outlined in Sect.~\ref{sec:nuc}. Lithium 
   production through the $\nu$-process in massive stars was also implemented 
   in one case (pink curves in Fig.~\ref{fig:SN}).

   The evolution of the thick-disc component is shown in the left panel, where 
   the x-axis reads, as usual, as a temporal axis. Because of the assumed fast 
   evolution \citep[see][]{2017MNRAS.472.3637G}, coupled to negligible 
   production of \element[ ][7]{Li} from both massive AGB stars 
   \citep{2000A&A...363..605V,2020A&A...641A.103V} and GCRs 
   \citep{2001A&A...374..646R,2012A&A...542A..67P} at the relevant 
   metallicities, the abundance of \element[ ][7]{Li} is predicted to remain 
   almost flat during most of the thick-disc evolution \cite[see 
     also][]{2016A&A...595A..18G,2019MNRAS.489.3539G}. Actually, owing to the 
   relatively high star formation efficiency, a mild decrease is expected due 
   to stellar astration, until novae start releasing large amounts of 
   \element[ ][7]{Li} through their outbursts, that is, $\sim$1--2~Gyr from the 
   onset of star formation \citep[see][their Fig.~3]{1999A&A...352..117R}. This 
   decrease is prevented if another short-lived (but still controversial) 
   \element[ ][7]{Li} factory is considered. The upper (pink) curve that does 
   not bend downwards for $-1 \le$~[Fe/H]/dex~$\le -0.5$ refers to a model 
   implementing the contribution from core-collapse SNe that produce 
   \element[ ][7]{Li} in neutrino-induced reactions in the He and C shells 
   \citep{1978Ap&SS..58..273D,1990ApJ...356..272W}. Starting from a primordial 
   high \element[ ][7]{Li} abundance, in accordance with SBBN predictions, and 
   considering all possible \element[ ][7]{Li} sources, the model reproduces 
   reasonably well the upper envelope of the GES iDR6 \element[ ][7]{Li} 
   measurements for thick-disc members. On the contrary, a model starting from 
   the Spite plateau value cannot explain the \element[ ][7]{Li} abundances 
   measured in thick-disc stars in the metallicity range $-0.8 
   \le$~[Fe/H]/dex~$\le -0.3$. It should be noted, though, that candidate 
   thick-disc stars were selected via chemical selection criteria (see 
   Sect.~\ref{sec:trace}). We will come back to this issue in 
   Sect.~\ref{sec:dyn}. It is also worth noticing that, although the model 
   attains ISM $A$(Li) values as high as $\sim$3.0~dex, very few stars form 
   with these high \element[ ][7]{Li} abundances, because of the low-level star 
   formation activity that characterises the thick disc during its latest 
   evolutionary stages.

   The middle panel of Fig.~\ref{fig:SN} is analogous to the left panel, but 
   refers to the thin-disc component. The blue circles in this plot are stars 
   that were assigned to the thin-disc component after a chemical selection 
   based on their metallicities and [$\alpha$/Fe] ratios. Metal-rich 
   $\alpha$-rich stars and super metal-rich solar-$\alpha$ stars found in the 
   solar vicinity today (magenta and light blue circles, respectively) are 
   likely migrating from inner Galactic regions \citep[][and references 
     therein]{2019A&A...623A..99G} and, as such, reflect different chemical 
   enrichment paths. Therefore, one should not compare their abundances to the 
   predictions of solar neighbourhood models \citep[see 
     also][]{2019A&A...623A..99G}. At variance with the results for the 
   thick-disc component, none of our theoretical curves for the local thin disc 
   is found to fold downwards in the $A$(Li)--[Fe/H] plane. This is due to the 
   less efficient star formation assumed for the thin disc 
   \citep[see][]{2017MNRAS.472.3637G} coupled to the continuous infall of gas 
   of primordial chemical composition that replenishes the ISM with fresh 
   \element[ ][7]{Li} on long timescales. Owing to the less efficient star 
   formation and longer infall timescale, the `reverse knee' indicating the 
   onset of significant \element[ ][7]{Li} production from novae in the lithium 
   tracks occurs at lower metallicities than in the thick disc, around 
   [Fe/H]~$\simeq -0.7$~dex or even lower, depending on the value assumed for 
   $A$(Li)$_\mathrm{P}$. Again, the upper envelope of the observations is best 
   explained by assuming $A$(Li)$_\mathrm{P}^{\mathrm{th}} \simeq 2.7$ dex.

   \begin{figure*}[ht]
     \centering
     \includegraphics[width=14cm]{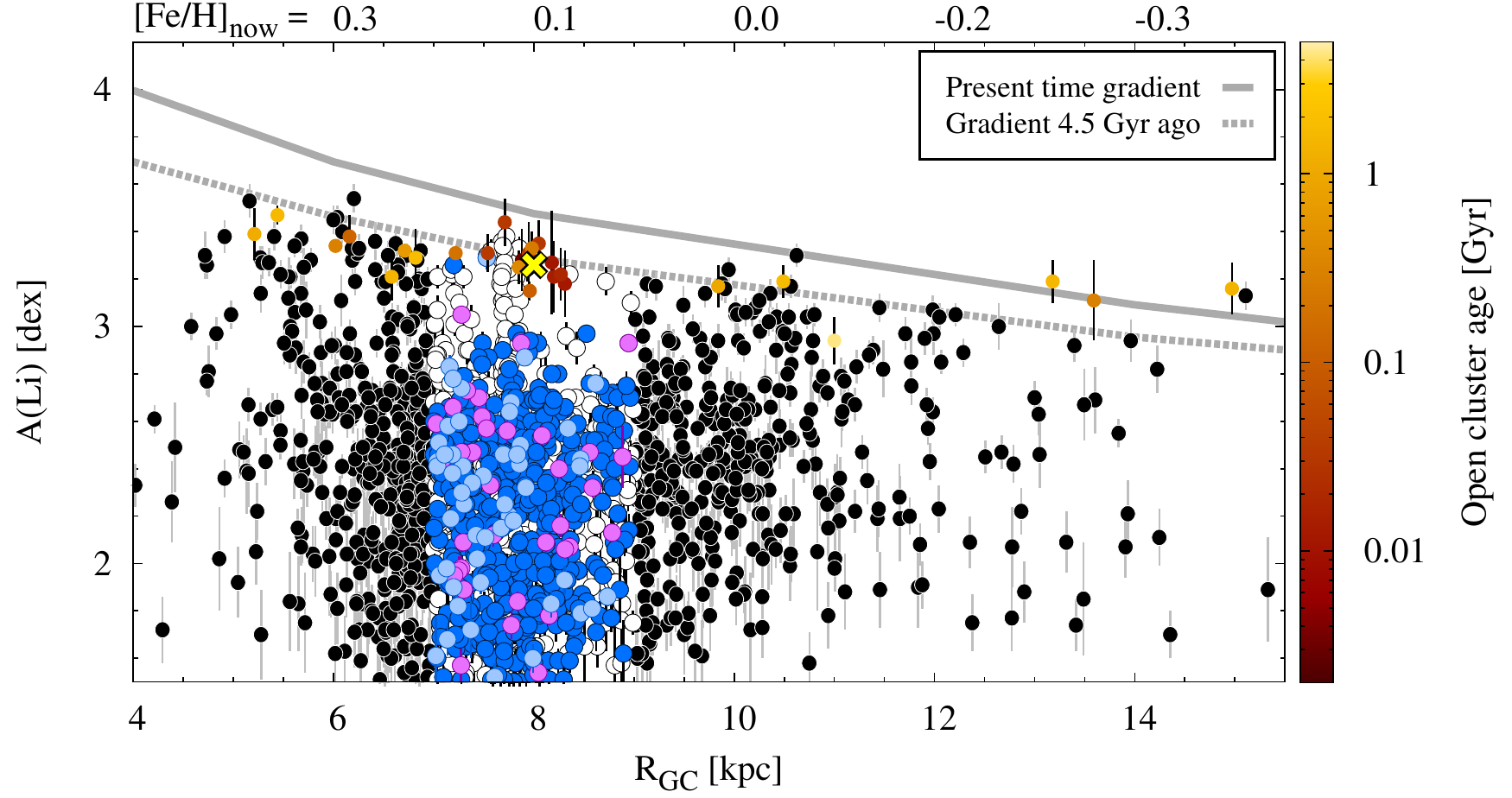}
          \caption{ Radial profiles of lithium abundance. The theoretical 
            predictions refer to the gradient at the Sun's birth ($t = 
            9.2$~Gyr, dotted curve) and at the current time ($t = 13.7$~Gyr, 
            solid line). Data for our sample of GES iDR6 field stars, 
            chemically separated into different sub-populations only in the 
            solar neighbourhood region, are shown in different colours. Blue, 
            magenta, and light blue circles identify, respectively, 
            low-$\alpha$, metal-rich $\alpha$-rich, and metal-rich 
            solar-$\alpha$ stars. Black circles refer to stars that inhabit 
            either the inner or the outer disc. Average maximum 
            \element[ ][7]{Li} abundances for 26 OCs selected from GES iDR6 are 
            also displayed (small colour-coded circles). The meteoritic 
            abundance from \citet{2009LanB...4B..712L} is indicated with the 
            multiplication sign. On the top x-axis we report the current 
            metallicity of the ISM as predicted by the model at different 
            radii.}
     \label{fig:GRAD}
   \end{figure*}

   In recent years, high-resolution spectroscopic surveys have identified a 
   decreasing trend of the lithium content in local field dwarfs at super-solar 
   metallicities \citep{2015A&A...576A..69D,2016A&A...595A..18G,
     2018A&A...615A.151B,2018A&A...610A..38F}. Such a puzzling behaviour has 
   called for some ad hoc reduction of the stellar yields or contamination from 
   thick-disc stars \citep{2017A&A...606A.132P}. Alternatively, a drop in the 
   nova outburst rate at high metallicities, which would lower the total 
   lithium production from these binary systems \citep{2018A&A...610A..38F,
     2019MNRAS.489.3539G}, or radial migration, bringing relatively old and 
   cool metal-rich thin-disc stars that have largely depleted their original 
   \element[ ][7]{Li} from inner radii to the solar neighbourhood 
   \citep{2019A&A...623A..99G}, have been suggested. Regarding the last point, 
   we note that, indeed, the vast majority of the stars labelled as super 
   metal-rich, solar-$\alpha$ stars in our sample are older than 3--4~Gyr and 
   cooler than 6000~K (light blue circles in Fig.~\ref{fig:SN}, right panel, 
   and Fig.~\ref{fig:ALiTeff}, respectively), which makes them highly 
   susceptible to internal lithium destruction processes.

   \citet{2020A&A...640L...1R}, based on GES iDR5, have shown that the average 
   maximum \element[ ][7]{Li} abundances of young, metal-rich OCs in the solar 
   neighbourhood tend to cluster around the meteoritic value; the most 
   metal-rich clusters found in the inner Galaxy display even higher 
   abundances, $A$(Li)~$> 3.4$~dex. We confirm those findings, which in turn 
   support the conclusions of \citet{2018AJ....155..138A} and 
   \citet{2019A&A...623A..99G}: the observed decrease in $A$(Li) for 
   super-solar metallicities is not real, but caused by sample selection 
   effects \citep[see also][]{2021A&A...649L..10C}. The lithium abundance in 
   the local ISM traced by young OCs and warm field dwarfs in our sample is not 
   decreasing; rather, it seems to flatten. However, we cannot exclude an 
   increase in the last 4.5~Gyr, due to possible effects of atomic diffusion on 
   warm, metal-rich F-type dwarfs that call for some caution when interpreting 
   the metal-rich data \citep{2021A&A...649L..10C}.

   From the middle panel of Fig.~\ref{fig:SN} it appears also that the GCE 
   model underestimates the \element[ ][7]{Li} content of the ISM in the $-0.5 
   \le$~[Fe/H]/dex~$< 0.0$ metallicity range. The culprit for this mismatch may 
   be the simplifying hypotheses made to implement nova nucleosynthesis in the 
   GCE code that lead to a monotonic increase in the \element[ ][7]{Li} 
   abundance in time since the first novae start to pollute the ISM. A 
   reduction of the nova outburst rate at high metallicities, as suggested by 
   \citet{2018A&A...610A..38F} and \citet{2019MNRAS.489.3539G}, may be needed 
   \citep[and is supported by independent observational evidence; 
     see][]{2014ApJ...788L..37G,2017MNRAS.469L..68G,2015ApJ...803...13Y}. A 
   larger \element[ ][7]{Li} production (potentially from novae) in the 
   metallicity range $-0.5 \le$~[Fe/H]~$< 0.0$ is, instead, necessary to fit 
   the data. In order to better assess these points, we analyse the 
   implications of a radial \element[ ][7]{Li} abundance gradient in the next 
   section.

   \subsection{The Galactic lithium gradient}
   \label{sec:grad}

   We followed the evolution of the abundance of \element[ ][7]{Li} in the ISM 
   at different positions across the thin disc by using the prescriptions of 
   model~1\,IM\,B by \citet{2018MNRAS.481.2570G}. This model assumes the 
   inside-out scenario for disc formation \cite[see][and references 
     therein]{2001ApJ...554.1044C} and a variable star formation efficiency 
   along the disc. We refer the reader interested in details to the original 
   publications. Here it will suffice to say that the model has been carefully 
   calibrated against a set of tight observational constraints\footnote{These 
     include the present-time gradients of several chemical elements along the 
     disc and their evolution, as well as the radial profiles of gas, stars, 
     and star formation rate \citep[see][for a detailed 
       discussion]{2001ApJ...554.1044C}.}, and that we did not make any attempt 
   to fine-tune the model parameters in order to improve the agreement with the 
   \element[ ][7]{Li} data. The ages of the OCs were taken from \citet[][see 
     Table~\ref{tab:OCsample}]{2020A&A...640A...1C}, while those of the field 
   stars were estimated using the aussieq2 tool \citep[see 
     Sect.~\ref{sec:Liabund} and][]{2020A&A...639A.127C}. The galactocentric 
   distances were homogeneously computed using {\fontfamily{cmtt} \selectfont 
     astropy} (see Sect.~\ref{sec:kindyn}).

   \begin{figure*}[ht]
     \centering
     \includegraphics[width=6.9cm]{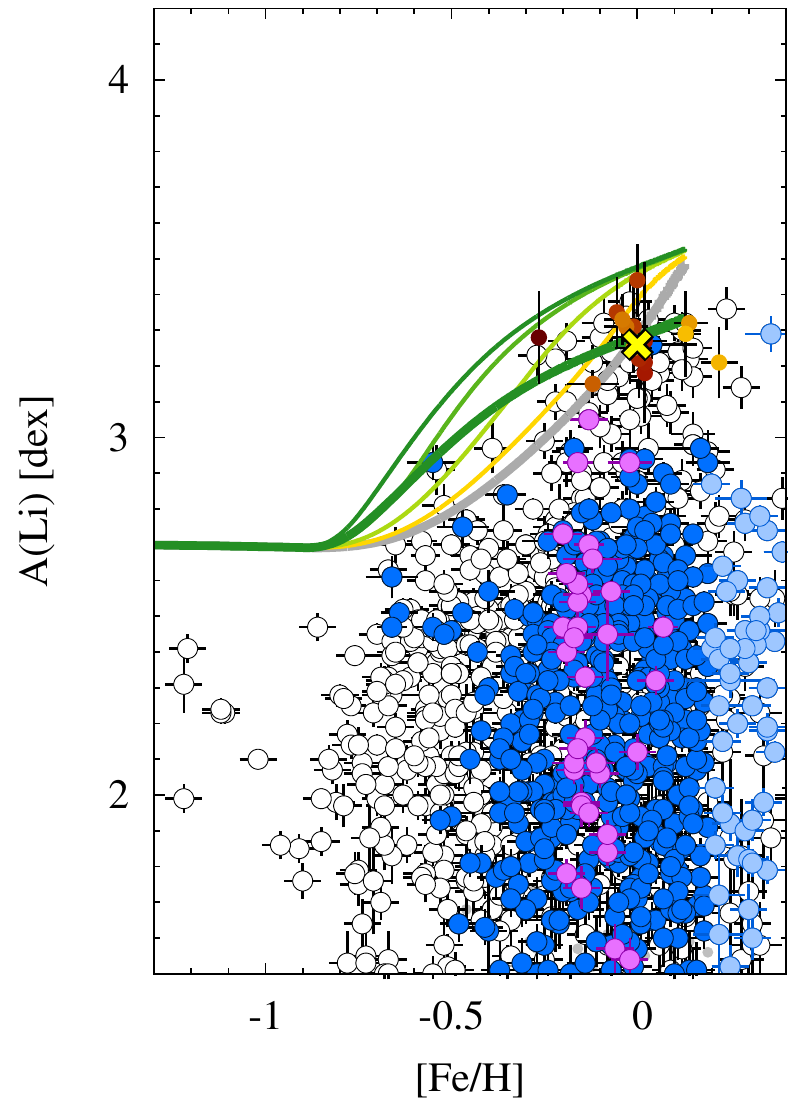}
     \includegraphics[width=6.9cm]{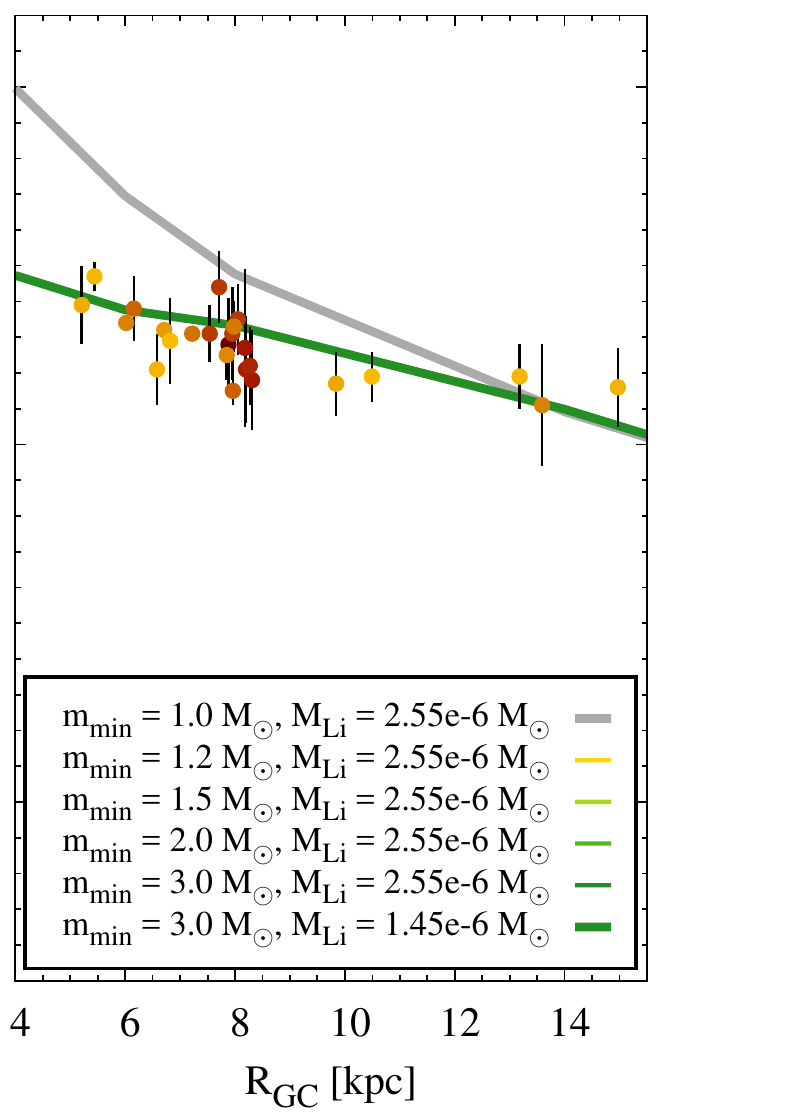}
          \caption{ Behaviour of $A$(Li) as a function of [Fe/H] in the local 
            thin disc ($R_\mathrm{GC} =$~7--9~kpc, \emph{left panel}) and as a 
            function of the galactocentric distance at the present time 
            \emph{(right panel)}. Lithium measurements for field low-$\alpha$, 
            mr$\alpha$r, and super metal-rich solar-$\alpha$ stars from GES 
            iDR6 are shown (blue, magenta, and light blue circles, 
            respectively, \emph{left panel} only) together with data for OCs 
            colour-coded according to the cluster's age (see the vertical bar 
            on the right in Fig.~\ref{fig:GRAD}). Only clusters with 
            ages~$<$~2~Gyr are shown. The predictions of our fiducial GCE model 
            are shown as thick grey lines. The thin solid lines illustrate the 
            effects of reducing the mass range of primary stars entering the 
            formation of nova systems (\emph{left panel} only) while keeping 
            the predicted current nova outburst rate fixed. The thick green 
            lines refer to a model in which the nova yield (i.e. the total 
            amount of \element[ ][7]{Li} ejected by a typical nova during its 
            lifetime, dubbed $M_\mathrm{Li}$ in the legend) is also reduced. See 
            text for details.}
     \label{fig:Mmin}
   \end{figure*}

   The theoretical gradients at different ages (now and 4.5~Gyr ago) are 
   compared to the data in Fig.~\ref{fig:GRAD}. The model used to compute the 
   gradients starts from the SBBN-predicted primordial \element[ ][7]{Li} 
   abundance and does not take lithium production from the $\nu$-process into 
   account. Unsurprisingly, the large majority of the field stars lie well 
   below the model predictions. This can be readily understood as an 
   age-temperature-rotation effect: only a minority of the stars in our sample, 
   in fact, are relatively young (ages~$<$ 2~Gyr), have $T_{\mathrm{eff}}$ in 
   excess of 6800~K and display high rotation rates (see 
   Figs.~\ref{fig:AgeTeff} and \ref{fig:ALivsini}), which are the right 
   characteristics to possibly show up as un-depleted stars \cite[see 
     Sect.~\ref{sec:trace}; see also][]{2020MNRAS.497L..30G,
     2021A&A...649L..10C}. The OCs span a much narrower age range, from 3~Myr 
   to about 2~Gyr, with only one cluster (NGC\,2243, at $R_{\mathrm{GC}} = 
   11$~kpc) being older than 4~Gyr. We retained this cluster to increase the 
   statistics in the outer disc. Its average maximum lithium abundance was 
   computed using 7 post turn-off members with $T_{\mathrm{eff}}$ between 6064~K 
   and 6314~K, in an effort to minimise the effects of \element[ ][7]{Li} 
   depletion. This notwithstanding, NGC\,2243 shows a \element[ ][7]{Li} 
   abundance $\sim$0.2--0.25 dex lower than those of the other outer disc 
   clusters. While the low \element[ ][7]{Li} content of NGC\,2243 could be 
   partly due to depletion as its stars are ageing, it could also be partly due 
   to a lower initial \element[ ][7]{Li} abundance inherited from the ISM when 
   the cluster formed 4.4~Gyr ago. Indeed, the GCE model predicts an increase 
   of 0.3--0.1~dex in the ISM \element[ ][7]{Li} abundance across the disc in 
   the last 4.5~Gyr (cfr. the dashed and solid lines in Fig.~\ref{fig:GRAD}). 
   The metallicity of NGC\,2243, [Fe/H]~$= -0.44 \pm 0.05$ dex, is lower than 
   the solar one and lower (by $\sim$0.2--0.4 dex) than those of the other 
   outer disc clusters, which would exclude an inner Galaxy origin and favour 
   the evolutive picture.

   For $R_\mathrm{GC} \le 10$~kpc, OCs and warm field dwarfs trace the same 
   \element[ ][7]{Li} abundance gradient. At larger galactocentric distances, 
   the gradient seems to flatten according to the OC data, while $A$(Li) keeps 
   decreasing in field stars, though it should be noted that the statistics in 
   the outer disc are very poor. Taken at face value, the data seem to suggest 
   that the observed gradient is flatter than the one predicted by our fiducial 
   model. In particular, the data for the inner Galaxy ($R_\mathrm{GC} \le 
   8$~kpc) agree very well with the predictions of the model at 4.5~Gyr ago. A 
   possible explanation is that high-metallicity clusters and field stars in 
   the inner disc were born with a higher lithium abundance, perhaps not too 
   far from that predicted by the model at the present time, 
   $A$(Li)~$\simeq$~3.5--3.8~dex, and show superficial abundances altered by 
   depletion or atomic diffusion \citep[see][]{2021A&A...649L..10C}. 
   Alternatively, we are seeing unaltered \element[ ][7]{Li} abundances and 
   the resulting mild inner disc gradient points to the need for a reduction of 
   lithium production at super-solar metallicities. This, in the framework of 
   our model, can be interpreted as evidence in favour of the hypothesis that 
   the nova outburst rate is reduced at high metallicities \citep[][and 
     references therein]{2018A&A...610A..38F,2019MNRAS.489.3539G}. On the other 
   hand, the predicted present-time gradient agrees very well with the OC data 
   in the outer disc. These external regions have not reached solar 
   metallicities yet and, as such, they have not entered the regime in which a 
   downward correction of the nova outburst rate could be required. Thus, a 
   model implementing a metallicity dependence of the nova progenitor formation 
   rate, as proposed by \citet{2019MNRAS.489.3539G}, offers a viable 
   explanation for the whole set of homogeneous \element[ ][7]{Li} abundances 
   examined in this paper. Another possible explanation is considered in the 
   following paragraphs.

  \subsection{An alternative model}
  \label{sec:change}

   \begin{figure*}[ht]
     \centering
     \includegraphics[width=14cm]{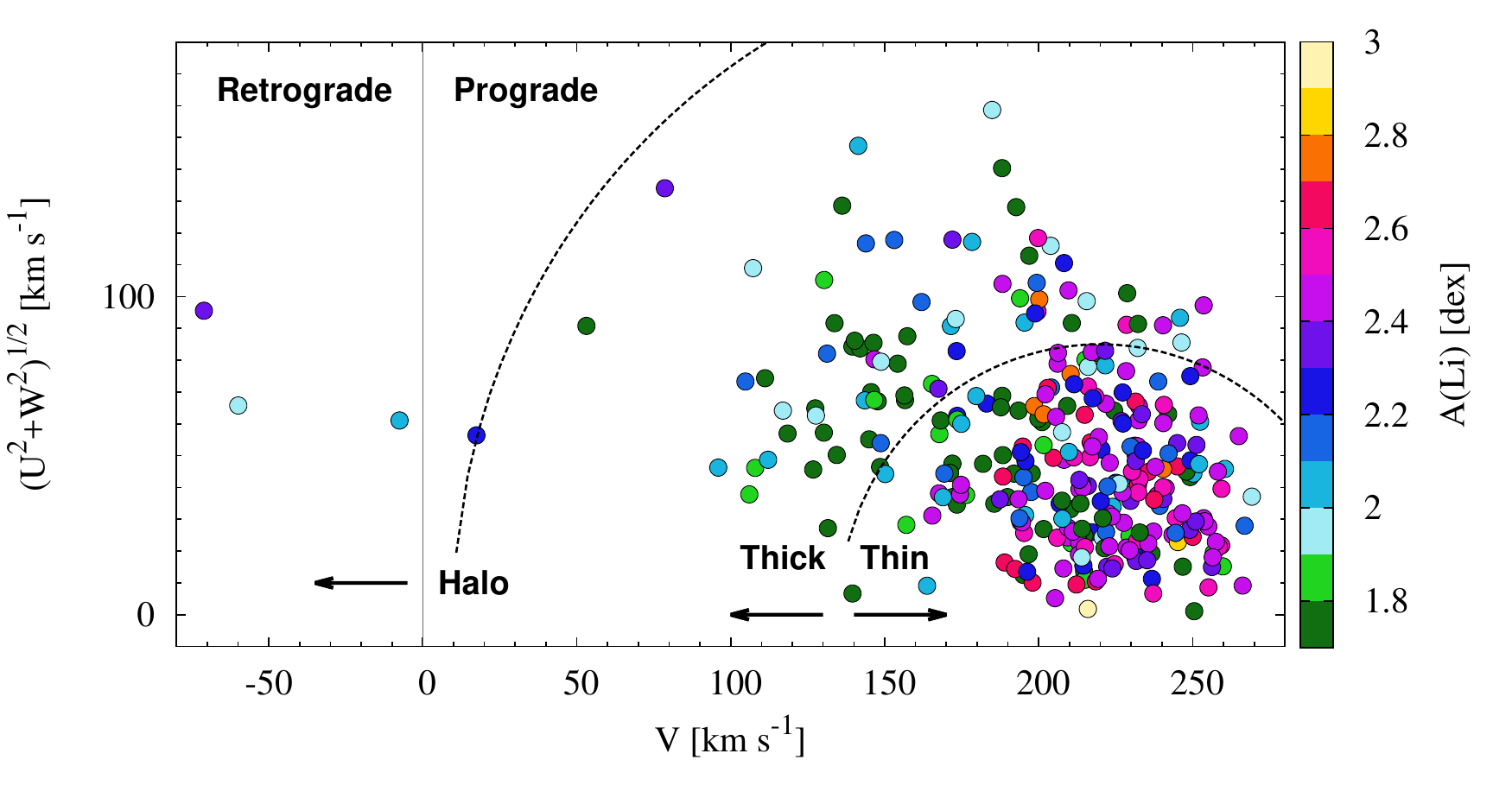}
          \caption{ Toomre diagram (LSR-corrected) for our sub-sample of 
            high-$\alpha$ stars, colour-coded according to their 
            \element[ ][7]{Li} abundance (see vertical bar on the right). 
            Constant total space velocities with respect to the LSR of 
            $V_{\mathrm{tot}} =85$ and 210~km~s$^{-1}$ are marked with dashed 
            curves. Low-$\alpha$, metal-rich $\alpha$-rich, and super 
            metal-rich solar-$\alpha$ stars (not shown to avoid overcrowding) 
            are consistent with thin-disc kinematics.}
     \label{fig:TOOMRE}
   \end{figure*}

   In our fiducial model, the star formation rate in the solar vicinity 
   declines slowly in time after an early period of intense activity. The nova 
   outburst rate, however, keeps always increasing mildly in time 
   \citep[see][their Fig.~3]{1999A&A...352..117R}, due to the contribution of 
   the numerous lowest-mass stars ($m \sim 1$~M$_\sun$) that enter the 
   formation of nova systems with lifetimes of the order of the age of the 
   Universe \citep[see e.g.][their Fig.~3]{2005A&A...430..491R}. If the 
   impact of such long-lived stars is lowered, one can expect a nova outburst 
   rate tracking more closely the star formation rate and, hence, a lower 
   \element[ ][7]{Li} production at late times.

   In Fig.~\ref{fig:Mmin}, left panel, we show the effects of reducing the 
   mass range of the primary stars that enter the formation of nova systems on 
   the $A$(Li) versus [Fe/H] trend predicted for the local thin disc. The 
   fiducial model, where the minimum mass of the stars entering nova systems is 
   set to 1~M$_\sun$, predicts the steepest increase in the lithium abundance 
   from the SBBN-predicted value to the meteoritic one. As the minimum mass 
   increases, the rise off the plateau is found to occur at lower metallicities 
   and the late evolution is characterised by flatter slopes (thin lines from 
   yellow to dark green in Fig.~\ref{fig:Mmin}, left panel). 

   All the models are calibrated to obtain the same current nova outburst rate 
   \citep[i.e. 17 events yr$^{-1}$;][and references 
     therein]{2015ApJ...808L..14I}. This makes the revised models overestimate 
   the \element[ ][7]{Li} abundance observed in meteorites. The thick green 
   line in Fig.~\ref{fig:Mmin}, left panel, shows the predictions of a model 
   (hereinafter dubbed alternative model) where both the mass range of WD 
   progenitors entering the formation of nova systems and the average 
   \element[ ][7]{Li} yield per nova are reduced so as to optimally fit the 
   local data within the errors. Of course, a strong degeneracy affects the 
   relevant model parameters and the one proposed here is not a unique choice. 
   A thorough investigation of the parameter space, however, is beyond the 
   scope of the present study and is deferred to a future paper.

   The right panel of Fig.~\ref{fig:Mmin} displays the present-day gradients of 
   $A$(Li) across the Milky Way disc predicted by the fiducial and alternative 
   models (grey and green thick lines, respectively), in comparison to our OC 
   data excluding the sole cluster with age~$>$~2~Gyr (NGC\,2243). Lacking the 
   contribution of the long-lived stars in the 1--3~M$_\sun$ mass range, the 
   alternative model predicts a gradient much flatter than that expected 
   according to the fiducial model. This milder theoretical gradient agrees 
   very well with the observed one traced by the average maximum 
   \element[ ][7]{Li} abundances of the selected OCs (see 
   Sect.~\ref{sec:Liabund}), though one should always keep in mind the caveats 
   spelled out in the previous sections. In particular: (i) the OC data include 
   stars on the cool side of the Li dip, thus, it is possible that these have 
   suffered some pre-main-sequence or main-sequence depletion (in which case 
   the depletion is metallicity dependent); (ii) higher metallicity means a 
   thicker sub-photospheric convection zone, hence, the OCs nearer the Galactic 
   centre, being more metal-rich, may have depleted their Li from a higher 
   level than the more metal-poor clusters further out in the disc and this 
   would have the effect of flattening the initial gradient; (iii) the effects 
   of atomic diffusion on the shape of the inner gradient still need to be 
   clarified. Therefore, at present we cannot discard a steeper gradient -- any 
   lying in between the green and the grey curves in Fig.~\ref{fig:Mmin}, right 
   panel.

   \subsection{Sanity check on membership and accreted stars}
   \label{sec:dyn}

   In Sect.~\ref{sec:trace} we adopt the chemical selection criteria defined by 
   \citet{2014A&A...567A...5R} and used in lithium studies by 
   \citet{2016A&A...595A..18G,2019A&A...623A..99G} and 
   \citet{2018A&A...610A..38F} as an option to separate thick-disc stars 
   (identified as high-$\alpha$ stars) from thin-disc ones (identified as 
   low-$\alpha$ stars). The adoption of different selection criteria, however, 
   leads to different samples of candidate thick- and thin-disc stars 
   \citep[e.g.][]{2003A&A...410..527B,2014A&A...562A..71B,2020ApJ...888...55F}. 
   In particular, with regard to \element[ ][7]{Li} evolution, 
   \citet{2018A&A...615A.151B} caution that the chemical selection may cause a 
   contamination of the thick-disc sample with thin-disc members, especially at 
   higher metallicities, which may lead to a spurious trend of increasing 
   \element[ ][7]{Li} abundance with metallicity for the thick disc. 

   In Fig.~\ref{fig:TOOMRE} we show the LSR-corrected (assuming $V_\mathrm{LSR} 
   = 220$ km s$^{-1}$) velocity components for our sub-sample of high-$\alpha$ 
   stars in a Toomre diagram. If a kinematical selection is applied, only 25 
   per cent of the high-$\alpha$ stars are classified as thick-disc members; 
   in particular, the stars with the highest \element[ ][7]{Li} abundances are 
   all thin-disc members \citep[in agreement with the results of][based on a 
     different sample of stars]{2018A&A...615A.151B}. Moreover, if a cut in age 
   is applied, so that only stars older than 8~Gyr are retained, no thick-disc 
   stars are left in the diagram with $A$(Li)~$>$ 2.5~dex.

\begin{figure}[t]
     \centering
     \includegraphics[width=7cm]{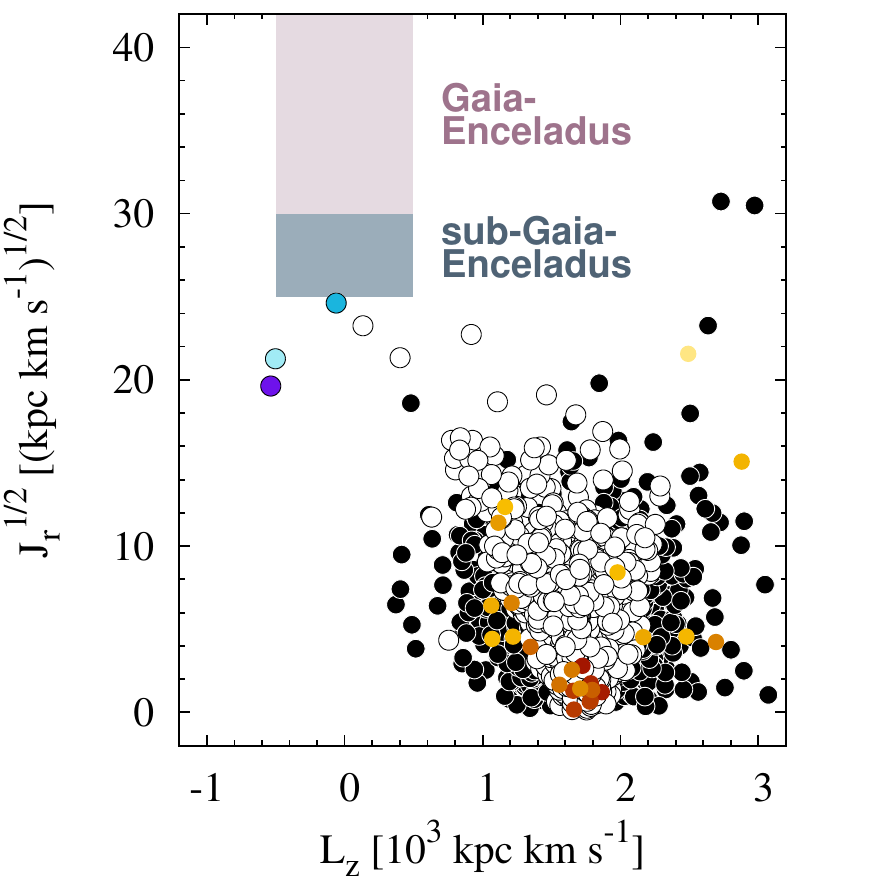}
          \caption{ $\sqrt{J_r}$ versus $L_z$ action space for our complete 
            sample. Empty circles are disc stars in the solar vicinity, full 
            black circles indicate targets observed at $R_{\mathrm{GC}} \le$ 
            7~kpc or $R_{\mathrm{GC}} \ge$ 9~kpc, and circles in shades from 
            yellow to maroon are OCs. The three retrograde halo stars are 
            colour-coded according to their \element[ ][7]{Li} content (see 
            Fig.~\ref{fig:TOOMRE}). The loci of {\it Gaia}-Enceladus and 
            sub-{\it Gaia}-Enceladus stars \citep[see][]{2020MNRAS.497..109F} 
            are highlighted.}
     \label{fig:FEUILLET}
   \end{figure}

   The velocity space is a useful tool to pinpoint the origin of stars, 
   especially when coherent moving groups can be identified \citep[see][for a 
     recent review]{2020ARA&A..58..205H}. We find that our sample includes 3 
   retrograde (counter-rotating) halo stars. We analyse their position in a 
   diagram of angular momentum $L_z$ versus square root of radial action in 
   Fig.~\ref{fig:FEUILLET}. \citet{2020MNRAS.497..109F} have pointed out that 
   {\it Gaia}-Enceladus stars are found in a restricted region of this space, 
   $30 \le \sqrt{J_r}$/(kpc km s$^{-1}$)~$\le 50$ and $-500 \le L_z$/(kpc km 
   s$^{-1}$)~$\le 500$ (clean sample); accreted stars may also be found at 
   slightly lower $\sqrt{J_r}$ values (sub-{\it Gaia}-Enceladus sample). One of 
   our retrograde halo stars is marginally compatible with being a member of 
   the latter sample. Clearly, a full chemical tagging may shed more light on 
   an accreted origin of our retrograde stars, but this is beyond the scope of 
   the present paper. We note, however, that all our retrograde stars have 
   $A$(Li) values consistent with the Spite plateau level \citep[see 
     also][]{2020MNRAS.496.2902M} and, hence, with no \element[ ][7]{Li} 
   enrichment as expected for stars hosted in dwarf galaxies according to 
   \citet{2021arXiv210411504M}. Those authors compute detailed chemical 
   evolution models for several dwarf spheroidal and ultra-faint dwarf galaxies 
   and show that, if most \element[ ][7]{Li} is forged in nova outbursts, as 
   also assumed in this paper, its abundance in Milky Way dwarf satellites must 
   lie flat at the Spite plateau level; only a few systems experiencing a more 
   prolonged star formation activity may host relatively metal-rich stars that 
   display higher lithium abundances.

   All our retrograde stars have $T_\mathrm{eff} < 6000$~K, namely, they fall on 
   the cool side of the dip. Measurements of \element[ ][7]{Li} abundances in 
   accreted stars on the warm side of the dip would be of the utmost importance 
   to characterise the evolution of \element[ ][7]{Li} in their galactic 
   progenitors, since they would possibly reflect un-depleted abundances. Such 
   stars must be rare -- because of stellar evolution effects combined with the 
   expected old ages -- but some relatively young objects gained through recent 
   mergers could still be found in large spectroscopic surveys targeting larger 
   stellar samples than GES, such as GALAH \citep{2015MNRAS.449.2604D}, or 
   WEAVE \citep{2012SPIE.8446E..0PD} and 4MOST \citep{2012SPIE.8446E..0TD} in 
   the future.

   \section{Discussion}
   \label{sec:discuss}

   Nowadays large spectroscopic surveys can secure precise radial velocities 
   and homogeneous abundances for light and heavy chemical elements, for as 
   many as hundreds of thousands of stars. This information, coupled to 
   accurate parallaxes and proper motions measured by the European Space Agency 
   {\it Gaia} mission, provides unprecedented constraints to GCE models. In 
   particular, when dealing with the Galactic evolution of lithium the 
   powerfulness of modern surveys is apparent. We can now select stars that 
   belong to different Galactic components and analyse how \element[ ][7]{Li} 
   enrichment proceeded in different substructures, including regions of the 
   disc located at different galactocentric distances to trace the gradient. We 
   can exclude contaminants, such as stars that migrated from the inner disc, 
   from local samples and get a cleaner picture of the evolutionary paths 
   followed by coherent groups of stars, rather than being dealing with the 
   superposition of different evolutionary sequences \citep[see][for a first 
     discussion on this aspect]{2019A&A...623A..99G}. As regards this point, it 
   is worth stressing that we identified inner disc intruders in our sub-sample 
   of solar neighbourhood stars via chemical selection criteria and that we 
   based our main conclusions on objects -- both clusters and field stars -- 
   that are younger than 2~Gyr and, hence, did not travel significant distances 
   across the disc \citep{2017A&A...597A..30A,2017MNRAS.470.4363C}.

   At present, the trends we see for \element[ ][7]{Li} abundance as a function 
   of metallicity, age, and galactocentric distance are still hampered by 
   several uncertainties. In most cases, \element[ ][7]{Li} abundances are 
   computed using LTE plane-parallel atmospheres; non-LTE corrections 
   calculated with 1D models are often applied, while 3D non-LTE corrections 
   are considered rarely. High rotation rates may broaden the lines and lead to 
   inaccurate results in case of automatic abundance derivation. The exact 
   location of the lithium dip in dependence of stellar parameters such as age 
   and metallicity is not well known and more OC studies are needed to sort out 
   all the pieces \citep[e.g.][and references therein]{2021AJ....161..159A,
     2021FrASS...8....6R}. Last in order of time, \citet{2021A&A...649L..10C} 
   warn that the effects of atomic diffusion in warm metal-rich stars deserve 
   further analysis before we can draw any firm conclusion on the actual 
   behaviour of $A$(Li) in the high-metallicity regime. This impacts both the 
   local late Galactic evolution and the slope of the gradient in the inner 
   disc.

   However, the \element[ ][7]{Li} abundance measured in meteorites provides a 
   good anchor for the models, as it has been `frozen' since the time of the 
   solar system formation 4.5 Gyr ago. Our GCE model, assuming that most 
   lithium is produced in nova outbursts, can explain the meteoritic lithium 
   abundance and, as we show in this paper, is able to reproduce either a 
   flatter or a steeper $A$(Li) gradient, depending on the assumed mass range 
   for the WD nova progenitors. Thus, determining the exact slope of the 
   gradient is a crucial step towards a better understanding of lithium 
   production on galactic scales.

   Another important issue that has to be fixed is that of the actual nova 
   outburst rate. Recently, \citet{2021ApJ...912...19D} have suggested that a 
   large population of highly obscured novae have been systematically missed in 
   optical searches. The current nova outburst rate would be $46.0 \pm 12.5$ 
   events yr$^{-1}$, larger than the estimate of 17 events yr$^{-1}$ assumed in 
   this work \citep[see also][]{2015ApJ...808L..14I}. This would point to a 
   lower average \element[ ][7]{Li} yield from novae, in better agreement with 
   the results of the majority of hydrodynamical simulations of the outburst.

   \section{Conclusions}
   \label{sec:end}

   Using spectroscopic data from the last internal data release of the GES and 
   astrometric information from {\it Gaia} EDR3, we have built a catalogue of 
   3210 stars in the Milky Way field that have (i) high-precision 1D LTE 
   \element[ ][7]{Li} abundances or upper limits, (ii) homogeneous stellar 
   parameters and (iii) [$\alpha$/Fe] ratios (for a sub-sample of stars), (iv) 
   stellar ages, (v) galactocentric distances, (vi) space velocity components, 
   and (vii) orbital parameters. The field star sample is complemented with 
   estimates of the average maximum \element[ ][7]{Li} abundances for 26 star 
   forming regions and OCs for which GES iDR6 delivers homogeneous stellar 
   parameters and abundances of confirmed members. All clusters, bar one, have 
   ages~$<$ 2~Gyr. After a careful inspection of the data, we conclude that:
   \begin{itemize}
     \item In the metallicity range $-0.3 <$~[Fe/H]/dex~$< +0.3$, where our GES 
       iDR6 field and OC samples overlap, field and cluster pre-main-sequence, 
       main-sequence, and turn-off stars trace the same evolution of lithium. 
       This is due to the fact that the last internal data release of the GES 
       includes field stars with ages in the range 1--2~Gyr that were absent in 
       previous releases of the survey \citep[see][]{2018MNRAS.473..185T}. 
       If only field stars older than 2~Gyr are considered, the average maximum 
       \element[ ][7]{Li} abundances estimated for the OCs lie systematically 
       above the upper envelope of the field star measurements. Field stars 
       older than 2~Gyr, in fact, fall in the lithium dip region or on the cool 
       side of the dip, where their \element[ ][7]{Li} abundances can be 
       affected by various degrees of depletion.
     \item The average maximum \element[ ][7]{Li} abundances of most OCs agree 
       well with the abundance of lithium measured in meteorites that have not 
       suffered any depletion.
     \item Our OC sample spans a wide range of galactocentric distances, 
       $5 \le R_\mathrm{GC}$/kpc~$< 15$. The current \element[ ][7]{Li} gradient 
       traced by OCs is a mild one, ranging from $A$(Li)~$\sim$ 3.4~dex in the 
       inner Galaxy to $A$(Li)~$\sim$ 3.1~dex in the outer disc. This gradient 
       is also supported by observations of warm field stars. It is unclear at 
       present whether atomic diffusion or other effects steepen the gradient 
       in the inner Galaxy.
     \item We have discovered three counter-rotating stars in our field star 
       sample that may be accreted stars. Their \element[ ][7]{Li} abundances 
       are consistent with the Spite plateau value, 
       $A$(Li)$_\mathrm{P}^\mathrm{obs} = 2.199 \pm 0.086$ 
       \citep{2010A&A...522A..26S}.
   \end{itemize}

   Recently, \citet{2019MNRAS.489.3539G} applied the parallel GCE model to the
   study of the evolution of \element[ ][7]{Li} in the Galactic thick and thin 
   discs. However, their analysis was limited to the solar vicinity. Here, we 
   take advantage of the homogeneous dataset secured by GES iDR6 for field and 
   OC stars to better assess the evolution of the abundance of 
   \element[ ][7]{Li} at high metallicities in the solar neighbourhood, as well 
   as its gradient across the Milky Way disc. We run a fiducial model, where 
   nearly 2 per cent of the stars with initial mass in the range 1--8~M$_\sun$ 
   enter the formation of nova systems as primary stars, and each nova is 
   assumed to eject $M^\mathrm{nova}_\mathrm{Li,\,tot} =$ 2.55~$\times 
   10^{-6}$~M$_\sun$ during its lifetime \citep{1999A&A...352..117R,
     2015ApJ...808L..14I}. We also run an alternate model, where 16 per cent of 
   the stars in the 3--8~M$_\sun$ mass range are primary stars in binary 
   systems that give rise to nova outbursts, and the average mass of 
   \element[ ][7]{Li} ejected in total by each nova is 
   $M^\mathrm{nova}_\mathrm{Li,\,tot} =$ 1.45~$\times 10^{-6}$~M$_\sun$. We 
   find that:
   \begin{itemize}
     \item The fiducial model underestimates the upper envelope of the 
       observations as traced by both warm field stars and OCs in the solar 
       neighbourhood, except for the super-solar metallicity regime, where the 
       \element[ ][7]{Li} abundances could be slightly overestimated. The 
       fiducial model also seems to predict a present-time gradient steeper 
       than the observed one. However, \citet{2021A&A...649L..10C} caution 
       that the possible effects of atomic diffusion on warm stars tracing the 
       high-metallicity trends have not yet been fully understood. These could 
       reconcile the model predictions with the observations.
       \item The alternative model fits (by construction) the upper envelope of 
         the local data in a $A$(Li)--[Fe/H] plane and the current gradient 
         along the disc.
       \item A model in which most \element[ ][7]{Li} comes from nova outbursts 
         explains the evolution of this fragile element well. However, the 
         parameters introduced to implement the nova nucleosynthesis are highly 
         degenerate and the proposed best-fitting model is, thus, not 
         necessarily unique.
   \end{itemize}
   More data are needed to probe if novae are the main lithium factories and to 
   assess whether their rate flattens out at late times. This latter 
   requirement, in particular, might have implications for the masses of the WD 
   nova progenitors, and it deserves further theoretical investigation. At 
   present, stellar yields of single low-, intermediate-, and high-mass stars 
   are too low for these stars to contribute sensibly to the evolution of 
   \element[ ][7]{Li} on galactic scales, apart perhaps from some contribution 
   from the $\nu$-process in massive stars that would counterbalance the 
   effects of stellar astration during early Galactic evolution. We also need 
   to better understand the mechanisms of lithium depletion in stars and their 
   dependences on stellar mass, metallicity, and other parameters, such as 
   rotation. There is a long way to go, but it is certainly worthwhile.

\begin{acknowledgements}
   This work is based on data products from observations made with ESO 
   Telescopes at the La Silla Paranal Observatory under programme IDs 
   188.B-3002, 193.B-0936, and 197.B-1074. These data products have been 
   processed by the Cambridge Astronomy Survey Unit (CASU) at the Institute of 
   Astronomy, University of Cambridge, and by the FLAMES/UVES reduction team 
   at INAF, Osservatorio Astrofisico di Arcetri. These data have been obtained 
   from the {\it Gaia}-ESO Survey Data Archive, prepared and hosted by the Wide 
   Field Astronomy Unit, Institute for Astronomy, University of Edinburgh, 
   which is funded by the UK Science and Technology Facilities Council. This 
   work was partly supported by the European Union FP7 programme through ERC 
   grant number 320360 and by the Leverhulme Trust through grant RPG-2012-541. 
   We acknowledge the support from INAF and from the Italian Ministry of 
   Education, University and Research (Ministero dell'Istruzione, 
   dell'Universit{\`a} e della Ricerca, MIUR) in the form of the grant Premiale 
   VLT 2012. The results presented here benefit from discussions held during 
   the Gaia-ESO workshops and conferences supported by the ESF (European 
   Science Foundation) through the GREAT Research Network Programme. This work 
   has made use of data from the European Space Agency (ESA) mission {\it Gaia} 
   (\url{https://www.cosmos.esa.int/gaia}), processed by the {\it Gaia} Data 
   Processing and Analysis Consortium (DPAC, 
   \url{https://www.cosmos.esa.int/web/gaia/dpac/consortium}). Funding for the 
   DPAC has been provided by national institutions, in particular the 
   institutions participating in the {\it Gaia} Multilateral Agreement. LM, SR, 
   GCas and AB acknowledges funding from MIUR Premiale 2016 MITiC. PB 
   acknowledges support from the French National Research Agency (ANR) funded 
   project ``Pristine'' (ANR-18-CE31-0017). VG acknowledges financial support 
   at SISSA from the European Social Fund operational Programme 2014/2020 of 
   the autonomous region Friuli Venezia Giulia. TB was funded by grant 
   No.~2018-04857 from The Swedish Research Council. AJK acknowledges support 
   from the Swedish National Space Agency (SNSA/Rymdstyrelsen). SLM is 
   supported by funding from the Australian Research Council via Discovery 
   Project DP180101791 and from the UNSW Scientia Fellowship program. FJE 
   acknowledges financial support from the Spanish MINECO/FEDER through the 
   grant AYA2017-84089 and MDM-2017-0737 at Centro de Astrobiolog{\'i}a 
   (CSIC-INTA), Unidad de Excelencia Mar{\'i}a de Maeztu, and from the European 
   Union's Horizon 2020 research and innovation programme under Grant Agreement 
   No.~824064 through the ESCAPE - The European Science Cluster of Astronomy 
   and Particle Physics ESFRI Research Infrastructures project.
\end{acknowledgements}


\bibliographystyle{aa} 
\bibliography{R21_lithium_BIB}

\end{document}